\documentclass[9pt,twocolumn,twoside]{pnas-new}
\templatetype{pnasresearcharticle} % Choose template 
\usepackage{amsmath}
\usepackage{amssymb}
\usepackage{bm}
\usepackage{accents}

\DeclareRobustCommand{\ubar}[1]{\underaccent{\bar}{#1}}

\definecolor{MyMaroon}{rgb}{0.247, 0.063, 0.22}
\definecolor{MyRed}{rgb}{0.412, 0.063, 0}
\definecolor{MyOrange}{rgb}{0.7, 0.275, 0.214}
\definecolor{MyYellow}{rgb}{0.7, 0.404, 0}
\definecolor{MyGreen}{rgb}{0.08, 0.486, 0}
\definecolor{MyTurquoise}{rgb}{0.003, 0.338, 0.324}
\definecolor{MyCyan}{rgb}{0.06, 0.634, 0.568}
\definecolor{MyBlue}{rgb}{0, 0.445, 0.7}
\definecolor{MyNavy}{rgb}{0, 0.211, 0.351}
\definecolor{MyGrey}{rgb}{0.4, .4, .4}

\title{Compressibility of complex networks}

\author[a,b]{Christopher W. Lynn}
\author[c,d,e,f,g,h,1]{Danielle S. Bassett} 

\affil[a]{Initiative for the Theoretical Sciences, Graduate Center, City University of New York, New York, NY 10016, USA}
\affil[b]{Joseph Henry Laboratories of Physics, Princeton University, Princeton, NJ 08544, USA}
\affil[c]{Department of Bioengineering, School of Engineering \& Applied Science, University of Pennsylvania, Philadelphia, PA 19104, USA}
\affil[d]{Department of Physics \& Astronomy, College of Arts \& Sciences, University of Pennsylvania, Philadelphia, PA 19104, USA}
\affil[f]{Department of Electrical \& Systems Engineering, School of Engineering \& Applied Science, University of Pennsylvania, Philadelphia, PA 19104, USA}
\affil[g]{Department of Neurology, Perelman School of Medicine, University of Pennsylvania, Philadelphia, PA 19104, USA}
\affil[h]{Department of Psychiatry, Perelman School of Medicine, University of Pennsylvania, Philadelphia, PA 19104, USA}
\affil[i]{Santa Fe Institute, Santa Fe, NM 87501, USA}

\leadauthor{Lynn} 

\significancestatement{Real-world networks are complex, comprising vast webs of interconnected elements performing a diverse array of social and biological functions. Common among many networks, however, is the pressure to be efficiently compressed -- either in the brain or in the genetic code. But just as files on a computer can be compressed to differing degrees, what makes one network more compressible than another? To answer this question, we adapt tools from information theory to quantify the compressibility of a network. Studying real-world and model networks, we find that hierarchical organization -- with tight clustering and heterogeneous degrees -- increases compressibility, enabling compressed representations across scales. Generally, our framework provides an information-theoretic method for investigating the interplay between network structure and compression.}

\authorcontributions{C.W.L. and D.S.B. conceived the project. C.W.L. designed the framework and performed the analyses with input from D.S.B. C.W.L. wrote the manuscript, and D.S.B. edited the manuscript.}
\authordeclaration{The authors declare no competing financial interests.}
\correspondingauthor{\textsuperscript{1}To whom correspondence should be addressed. E-mail: dsb@seas.upenn.edu }

\keywords{Information theory $|$ Complex networks $|$ Rate-distortion $|$ Compression} 

\begin{abstract}
Many complex networks depend upon biological entities for their preservation. Such entities, from human cognition to evolution, must first encode and then replicate those networks under marked resource constraints. Networks that survive are those that are amenable to constrained encoding, or, in other words, are \emph{compressible}. But how compressible is a network? And what features make one network more compressible than another? Here we answer these questions by modeling networks as information sources before compressing them using rate-distortion theory. Each network yields a unique rate-distortion curve, which specifies the minimal amount of information that remains at a given scale of description. A natural definition then emerges for the compressibility of a network: the amount of information that can be removed via compression, averaged across all scales. Analyzing an array of real and model networks, we demonstrate that compressibility increases with two common network properties: transitivity (or clustering) and degree heterogeneity. These results indicate that hierarchical organization -- which is characterized by modular structure and heterogeneous degrees -- facilitates compression in complex networks. Generally, our framework sheds light on the interplay between a network's structure and its capacity to be compressed, enabling investigations into the role of compression in shaping real-world networks.
\end{abstract}

\dates{This manuscript was compiled on \today}
\doi{\url{www.pnas.org/cgi/doi/10.1073/pnas.XXXXXXXXXX}}

\begin{document}

\maketitle
\thispagestyle{firststyle}
\ifthenelse{\boolean{shortarticle}}{\ifthenelse{\boolean{singlecolumn}}{\abscontentformatted}{\abscontent}}{}

% If your first paragraph (i.e. with the \dropcap) contains a list environment (quote, quotation, theorem, definition, enumerate, itemize...), the line after the list may have some extra indentation. If this is the case, add \parshape=0 to the end of the list environment.

\dropcap{C}omplex networks are often encoded in biology, and thereby utilized and replicated by biological systems. The brain encodes language \cite{Sizemore-01}, knowledge \cite{Vazquez-02}, music \cite{Liu-02}, social \cite{Girvan-01, Brush-01}, and transportation networks \cite{Kalakoski-01}; the human mind uses these internal representations to engage in linguistic communication, build on existing understanding, sing a victorious melody, strengthen a valuable friendship, and walk the covered holloways \cite{Lynn-08}. Similarly, biological networks among molecular and cellular components are encoded at various scales in genetic material \cite{Gavin-01, Lynn-05, Vertes-01, Whitaker-01}; and evolution uses these encodings to propagate network topologies in a surviving species. From brains to genes, the biological materials that encode complex networks operate under marked constraints on time, energy, metabolism, and physical extent, among others. Such constraints determine which networks persist into the future; in particular, those whose topology can be efficiently encoded. These shared constraints raise a fundamental question: How does the structure of a network facilitate efficient encodings?

Encoding a network (indeed, encoding any piece of information) involves a natural trade-off between simplicity and accuracy. One could construct a simple representation that omits the fine-scale details of a network. Or one could build a representation that captures a network's intricate structure, but is complicated and unwieldy. An efficient encoding strikes an optimal balance between simplicity and accuracy; that is, it is a compression \cite{Shannon-01, Cover-01}. In fact, compression -- a foundational branch of information theory -- has provided key insights into optimal network representations, yielding principled algorithms for constructing coarse-grained maps of complex systems \cite{Rosvall-02, Rosvall-03, Slonim-01}.

Building upon this progress, here we investigate how the structure of complex networks facilitates compression. Intuitively, just as natural images are easier to compress than white noise due to their visual patterns and regularities, so too should networks with strong structural regularities be more compressible than random networks. But do homogeneous topologies, such as those found in lattice-like networks, make systems more compressible, or is compression facilitated by the hierarchical organization found in many real networks? To answer these questions, here we develop a framework for quantifying the compressibility of complex networks. Applying our framework to several real and model networks, we identify specific network features that facilitate compression. Together, these results elucidate how a network's topology impacts its compressibility, and suggest that many real-world networks may be shaped by the pressure to be compressed.

\section*{Rate-distortion theory of network clustering}

In compression \cite{Cover-01}, one begins with an information source, a sequence of items that defines the object of interest. In the context of networks, the details of information flow often vary from one setting to another. Therefore, a logical choice for the information source is a random walk, which encodes only the information contained in a network's structure and nothing more \cite{Rosvall-02}. One then seeks to reduce the amount of information in the sequence by constructing a coarse-grained representation of the network. We remark that the \textit{lossless} compression of random walks has provided important information-theoretic perspectives on the problem of community detection \cite{Rosvall-02}. By contrast, rather than choosing a specific number of communities, here we are interested in analyzing the compressibility of networks across all scales. To do so, we employ rate-distortion theory, the foundation of \textit{lossy} compression. Importantly, rate-distortion theory will enable tractable strategies for compressing networks across all scales and, in doing so, will allow us to develop an intuitive definition for compressibility. 

\subsection*{Compressing random walks}

\begin{SCfigure*}[\sidecaptionrelwidth][t]
\centering
\includegraphics[width = 1.5\linewidth]{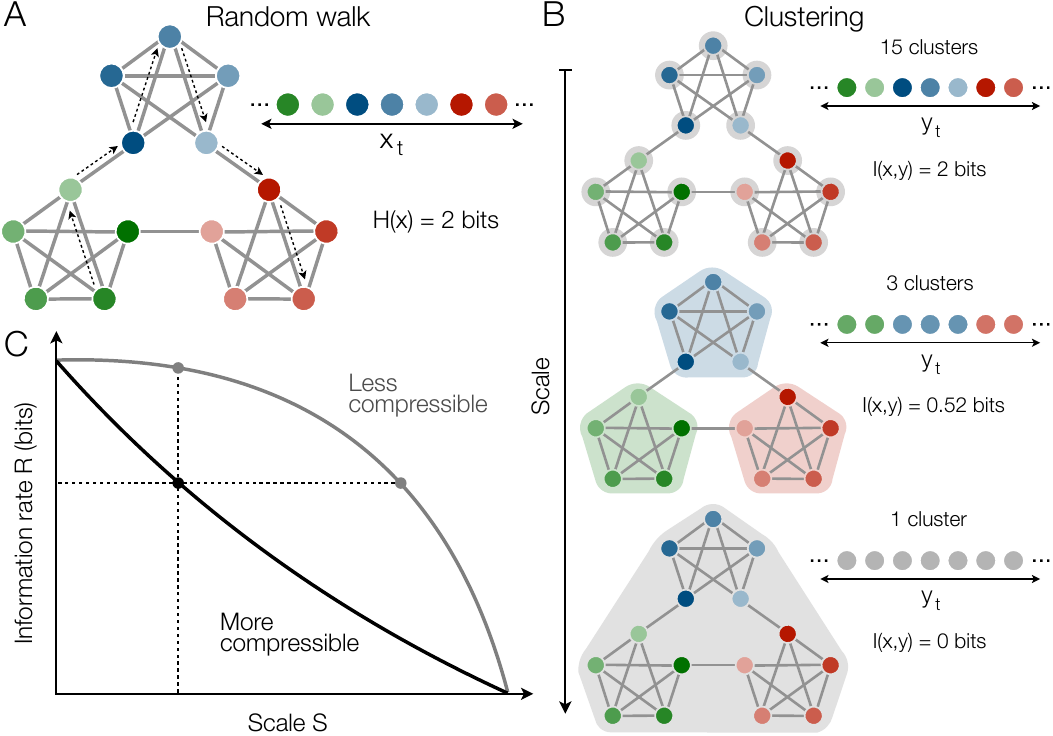}
\caption{\textbf{Rate-distortion theory of random walks on networks.} (\textit{A}) A simple network with $N=15$ nodes, each with constant degree $k = 4$. A random walk $\bm{x}$ generates information at a rate $H(\bm{x}) = 2$ bits. (\textit{B}) Network clusterings across various scales of description. For $n=15$ clusters, each containing its own node, the sequence communicates $I(\bm{x},\bm{y}) = H(\bm{x}) = 2$ bits of information (\textit{Top}). For $n=3$ clusters, each corresponding to one of the three modules in the original network, the information rate is $I(\bm{x},\bm{y}) = 0.52$ bits (\textit{Middle}). For $n=1$ cluster containing the entire network, the sequence no longer communicates information (\textit{Bottom}). (\textit{C}) Schematic of the optimal information rate $R$ as a function of the scale of description $S$ for networks that are either more compressible (black) or less compressible (grey). For more compressible networks, one can achieve a lower information rate at a given scale of description (vertical line), and one can achieve a more fine-grained description for a given information rate (horizontal line).}
\label{rate_distortion}
\end{SCfigure*}

\hspace{0pt} To see how compression unfolds in practice, consider the network in Fig. \ref{rate_distortion}\textit{A}. A random walk on the network defines a sequence of nodes $\bm{x} = (x_1,x_2,\hdots)$, with each node transitioning to one of its four neighbors uniformly at random. The rate at which this sequence generates information is given by the entropy $H(\bm{x})$, which (because there are four possible nodes at each step) equals $2$ bits (see Materials and Methods for a definition of $H(\bm{x})$). To reduce the amount of information in the sequence, we can construct a coarse-grained representation by clustering nodes together \cite{Rosvall-02, Rosvall-03, Slonim-01}. This clustering yields a new sequence $\bm{y} = (y_1,y_2,\hdots)$, where $y_t$ is the cluster containing node $x_t$ (Fig. \ref{rate_distortion}\textit{B}), which communicates information at a rate equal to the mutual information $I(\bm{x},\bm{y}) = H(\bm{y}) - H(\bm{y}|\bm{x})$ \cite{Shannon-01, Cover-01, Rosvall-03, Slonim-01}. If the clusters are chosen deterministically, as is common \cite{Girvan-01, Rosvall-02, Leskovec-02}, then the conditional entropy $H(\bm{y} | \bm{x})$ vanishes, and the information rate simplifies to the entropy of the clustered sequence, $I(\bm{x},\bm{y}) = H(\bm{y})$.

Consider, for example, a trivial clustering in which each node belongs to its own cluster (Fig. \ref{rate_distortion}\textit{B, top}). In this case, we maintain a complete description of the network, but we have not reduced the information rate, since $I(\bm{x},\bm{y}) = H(\bm{x}) = 2$ bits. By contrast, consider the opposite setting in which all nodes belong to the same large cluster (Fig. \ref{rate_distortion}\textit{B, bottom}). Now we have reduced the information rate to zero ($I(\bm{x},\bm{y}) = 0$ bits), but all details about the network structure have been lost. Between these two extremes lies a range of clusterings (such as that in Fig. \ref{rate_distortion}\textit{B, middle}), each inducing its own information rate and yielding a unique distortion of the network structure.

\subsection*{Scale as a measure of distortion}

\hspace{0pt} Building representations that strike an optimal balance between minimizing information rate while also minimizing distortion is precisely the purview of rate-distortion theory \cite{Shannon-01, Cover-01}. As in any rate-distortion problem, one must choose a specific definition for the distortion of the object of interest. When clustering a network, a natural choice for the distortion presents itself: the scale of description. Specifically, for a network with $N$ nodes and a clustering with $n$ clusters, we define the scale to be $S = 1 - \frac{n-1}{N}$. For example, if $n = N$, then we have an exact fine-grained description of the network at a scale $S = 1/N$ (Fig. \ref{rate_distortion}\textit{B, top}); whereas if $n=1$, then one cluster encloses the entire network and $S = 1$ (Fig. \ref{rate_distortion}\textit{B, bottom}).

At each scale $S$ (equivalently, for each number of clusters $n$), we seek to identify the clustering that minimizes the information rate $I(\bm{x},\bm{y})$. This optimal information rate, denoted $R(S)$, defines a unique rate-distortion curve for each network (Fig. \ref{rate_distortion}\textit{C}). If a network is easier to compress, then at each scale $S$ one should be able to find a clustering that is more efficient, reducing the information rate $R$ (Fig. \ref{rate_distortion}\textit{C, vertical line}); similarly, for a given information rate $R$ one should be able to construct a more fine-grained clustering, decreasing the scale $S$ (Fig. \ref{rate_distortion}\textit{C, horizontal line}). Thus, in order to quantify the compressibility of a network, we must first be able to compute its rate-distortion curve.

\section*{Computing the rate-distortion curve of a network}

Computing the rate-distortion curve $R(S)$ of a network -- in particular, doing so efficiently to enable applications to large systems -- poses two distinct challenges. First, we must estimate the mutual information $I(\bm{x},\bm{y})$ for different clusterings; and second, we must identify the clusterings that minimize this information rate across all scales.

Although estimating mutual information is generally difficult \cite{Archer-01}, the simplicity of our setup allows for tractable upper and lower bounds (see Materials and Methods). Of particular interest is the upper bound $\bar{I}(\bm{x},\bm{y}) \ge I(\bm{x},\bm{y})$, which follows by approximating the clustered sequence $\bm{y}$ as Markovian (a property that we note is not guaranteed, even though the original random walk $\bm{x}$ is Markovian \cite{Cover-01}). Rather than minimizing the information rate $I(\bm{x},\bm{y})$ directly, we instead minimize the upper bound $\bar{I}(\bm{x},\bm{y})$, yielding an upper bound $\bar{R}(S)$ on the rate-distortion curve. For simplicity, in what follows we often refer to $\bar{I}(\bm{x},\bm{y})$ as the information rate and $\bar{R}(S)$ as the rate-distortion curve.

To compute $\bar{R}(S)$ -- that is, to find clusterings that minimize the information rate $\bar{I}(\bm{x},\bm{y})$ -- we employ a greedy clustering algorithm. Beginning with $n = N$ clusters, each containing its own node, we combine the pair of clusters that yields the largest reduction in the information rate $\bar{I}(\bm{x},\bm{y})$. Repeating this agglomerative process across all scales $S$ (until only one cluster remains), we arrive at an estimate for the rate-distortion curve $\bar{R}(S)$. To speed up the calculation, rather than searching through all $n \choose 2$ pairs of clusters at each step, we only consider a limited number of pairs chosen via principled heuristics (see Materials and Methods). Importantly, these heuristics do not affect the definitions of information-theoretic quantities, such as the rate $I(\bm{x},\bm{y})$ and upper bound $\bar{I}(\bm{x},\bm{y})$. In practice, not only do these heuristics enable applications to networks of approximately $10^3$ nodes, they also improve the accuracy of the rate-distortion estimates themselves (see Supporting Fig. 1).

\begin{figure*}[t]
\centering
\includegraphics[width = .9\linewidth]{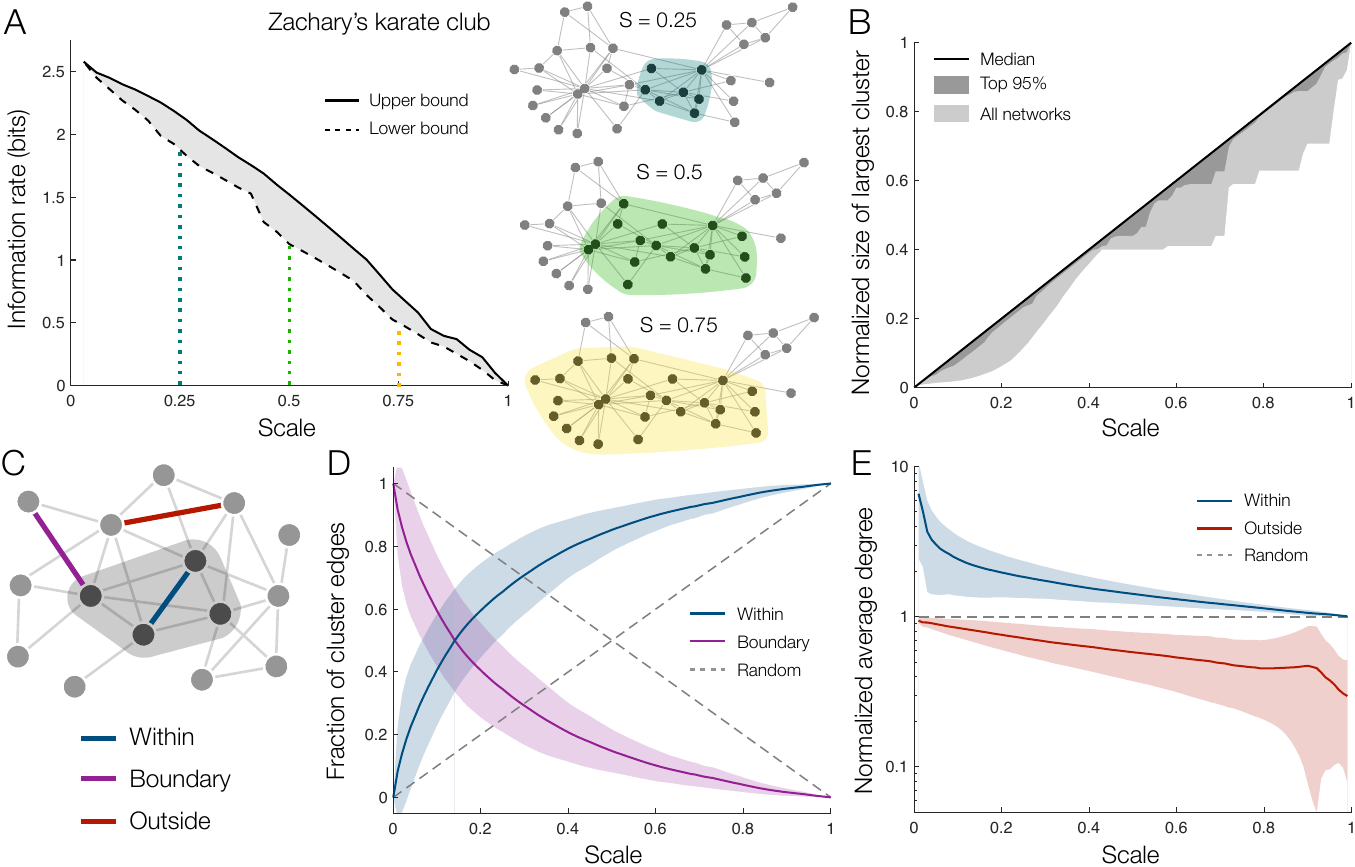}
\caption{\textbf{Properties of optimal clusterings.} (\textit{A}) Upper bound (solid line) and lower bound (dashed line) on the optimal information rate $R(S)$ as a function of the scale of description $S$ for Zachary's karate club network \cite{Zachary-01}. Across all scales, the optimal compression includes one large cluster, which we illustrate for $S = 0.25$, $0.5$, and $0.75$ (\textit{Right}). (\textit{B}) Size of the largest cluster in a compression, normalized by the size of the network $N$, as a function of the scale $S$ for the real networks in Supporting Table 1. The median over real networks (solid line) matches the largest possible normalized cluster size, $(N - n + 1)/N = S$, indicating that (across all scales) most networks admit one large cluster of maximal size. (\textit{C}) Illustration of edges within the one large cluster (blue), on the boundary of the cluster (purple), and outside the cluster (red). (\textit{D}) Fraction of the $k_c$ edges emanating from the large cluster that either connect to nodes outside the cluster $1 - G_{cc}/k_c$ (purple) or remain within the cluster $G_{cc}/k_c$ (blue) as a function of the scale $S$. (\textit{E}) Average degree of nodes inside (blue) and outside (red) the large cluster, normalized by the average degree of the network, as a function of the scale $S$. In panels (\textit{D}) and (\textit{E}), solid lines and shaded regions represent averages and one-standard-deviation error bars, respectively, over the real networks (Supporting Table 1), and dashed lines correspond to clusters with nodes selected at random.}
\label{clustering}
\end{figure*}

We are now prepared to compute the rate-distortion curve for a specific system. In Fig. \ref{clustering}\textit{A}, we plot the upper and lower bounds on the rate-distortion curve $R(S)$ for Zachary's karate club network \cite{Zachary-01}. As is true for all networks (see Materials and Methods), the two bounds are exact at both the minimum scale $S = 1/N$ (when the information rate simply equals the entropy of random walks $H(\bm{x})$) and the maximum scale $S = 1$ (when the information rate is zero). Moreover, the two bounds remain close across all intermediate scales (Fig. \ref{clustering}\textit{A}), demonstrating that the upper bound $\bar{R}(S)$ provides a good approximation to the true rate-distortion curve $R(S)$. To understand how the rate-distortion curve depends on the structure of a network, however, it helps to examine the properties of optimal compressions themselves.

\section*{Properties of optimal compressions}

Using the framework developed above, we are ultimately interested in studying compression in real systems. The networks chosen for analysis span from communication networks (including semantic, language, and music networks) and information networks (including hyperlinks on the web and citations in science) to social networks, animal and protein interactions, transportation networks, and structural and functional connections in the brain (see Materials and Methods; Supporting Table 1). Although these networks encompass a wide range of systems bridging several orders of magnitude in size, they are all encoded biologically, either in genetic material or in the neural code.

\subsection*{Emergence of one large cluster}

\hspace{0pt} To begin, we compute the rate-distortion curve $\bar{R}(S)$ for each of the above networks, and we confirm that these upper bounds provide good approximations to the true rate-distortion curves $R(S)$ (see Supporting Fig. 2). In the process of computing $\bar{R}(S)$, our compression algorithm also provides estimates for the optimal clusterings over all scales. Examining the structure of these compressions, we find a striking consistency across different networks. As can be observed in Zachary's karate club (Fig. \ref{clustering}\textit{A, Right}), rather than dividing the network into multiple clusters of moderate size, optimal compressions tend to comprise one large cluster containing $N - n + 1 = SN$ nodes and $n-1$ minimal clusters each containing one node. In fact, among the networks studied, this tendency to form one large cluster is a nearly ubiquitous feature of optimal compressions (Fig. \ref{clustering}\textit{B}).

We remark that the clustering that minimizes the information rate need not (and indeed, does not) provide a faithful characterization of a network's community structure, as is the goal in community detection \cite{Rosvall-02, Rosvall-03, Slonim-01}. Instead, we find that optimal compressions seek to identify the group of nodes that can be combined to maximally reduce the information rate. By dividing the network into two parts -- one inside the large cluster and the other outside -- the challenge of compressing random walks thus resembles the graph partitioning problem \cite{Bulucc-01}, which has generated key insights about the modular structure of networks across scales \cite{Leskovec-02}. This simplification, in turn, allows us to develop analytic predictions about the properties of optimal compressions and the structures of compressible networks.

\subsection*{Information rate of optimal compressions}

\hspace{0pt} Although our framework is general, applying to any weighted, directed network (see Materials and Methods), in order to make analytic progress, here we focus on the special case of an unweighted, undirected network with adjacency matrix $G_{ij}$. For such a network, the entropy of random walks takes the simple form $H(\bm{x}) = \frac{1}{2E}\sum_i k_i \log k_i$, where $k_i = \sum_j G_{ij}$ is the degree of node $i$, $E = \frac{1}{2}\sum_{ij} G_{ij}$ is the number of edges in the network, and $\log(\cdot)$ is base 2 such that information is measured in bits.

Now consider forming one large cluster $c$. One can show (see Materials and Methods) that the information rate of the clustered network is given by
\begin{align}
\label{Ibar}
\bar{I}(\bm{x},\bm{y}) = \frac{1}{2E} &\bigg[\sum_{i\not\in c} k_i \log k_i + k_c \log k_c  \\
& \quad -2\sum_{i\not\in c} G_{ic}\log G_{ic} - G_{cc}\log G_{cc}\bigg], \nonumber
\end{align}
where $k_c = \sum_{i\in c} k_i$ is the sum of the degrees of the nodes in $c$, $G_{ic} = \sum_{j\in c} G_{ij}$ is the number of edges connecting nodes in $c$ to a given node $i$, and $G_{cc} = \sum_{ij\in c} G_{ij}$ is the number of edges connecting nodes within $c$.

\subsection*{Information content of different edges}

\hspace{0pt} Using Eq. \ref{Ibar}, can we predict the properties of the optimal cluster $c$? More broadly, can we anticipate the types of network topologies that facilitate compression? To answer these questions, it helps to group the edges in a network into three distinct categories (Fig. \ref{clustering}\textit{C}): those connecting nodes \textit{within} $c$, those connecting nodes \textit{outside} of $c$, and those on the \textit{boundary} of $c$ (connecting nodes within $c$ to nodes outside of $c$). We can gauge which type of edge is preferred over the others by comparing their contributions to the information rate (Eq. \ref{Ibar}). An optimal compression will maximize the number of edges that are informationally preferred (contributing only weakly to the information rate), while limiting edges that are informationally costly.

For example, adding an edge within $c$ increases the information rate by $\Delta\bar{I}^{\text{within}} \approx \frac{1}{2E}(2\log k_c - 2\log G_{cc})$. By contrast, adding an edge on the boundary of $c$ (say, connecting $c$ to a node $i\not\in c$) yields an increase of roughly $\Delta \bar{I}^{\text{boundary}} \approx \frac{1}{2E}(\log k_i + \log k_c - 2\log G_{ic})$. For a large cluster $c$, we have $k_c, G_{cc} \gg k_i, G_{ic}$, from which one can show that $\Delta\bar{I}^{\text{within}} \lesssim \Delta \bar{I}^{\text{boundary}}$ (see Supporting Information). Thus, edges within the large cluster are informationally preferred to those on the boundary, suggesting that the large cluster will seek to combine groups of nodes that are tightly-connected to one another and sparsely connected to the rest of the network. Indeed, in real networks, we find that among the $k_c$ edges emanating from the large cluster, the proportion $1 - G_{cc}/k_c$ that connects to the rest of the network is much smaller than chance (Fig. \ref{clustering}\textit{D}). This proportion of edges leaving the cluster is a well-studied quantity, known as the conductance or Cheeger constant of a network \cite{Leskovec-02}. Thus, networks with low conductance -- such as those with modular structure and strong transitivity (the tendency for nodes to form triangles, also known as clustering) -- should be highly compressible \cite{Newman-05, Leskovec-02}. This is our first hypothesis about the impact of network structure on compressibility.

We now consider an edge connecting two nodes $i$ and $j$ outside of $c$, which increases the information rate by approximately $\Delta \bar{I}^{\text{outside}} \approx \frac{1}{2E}(\log k_i + \log k_j)$. As before, one can show that $\Delta\bar{I}^{\text{within}} \lesssim \Delta \bar{I}^{\text{outside}}$ (see Supporting Information), demonstrating that edges within the large cluster are informationally preferred to those outside the cluster. In turn, this preference for the large cluster to include as many edges as possible suggests that $c$ will favor high-degree nodes over low-degree nodes, which we confirm in real networks (Fig. \ref{clustering}\textit{E}). This result leads to our second hypothesis: networks should be more compressible if they have heterogeneous degrees (or heavy-tailed degree distributions), containing ``rich-clubs'' of high-degree hub nodes \cite{Benson-01, Barabasi-01}. Given the predictions that modular and heterogeneous topologies facilitate compression, we now propose a quantitative definition for the compressibility of a network.

\section*{Quantifying network compressibility}

\begin{figure*}[t]
\centering
\includegraphics[width = \linewidth]{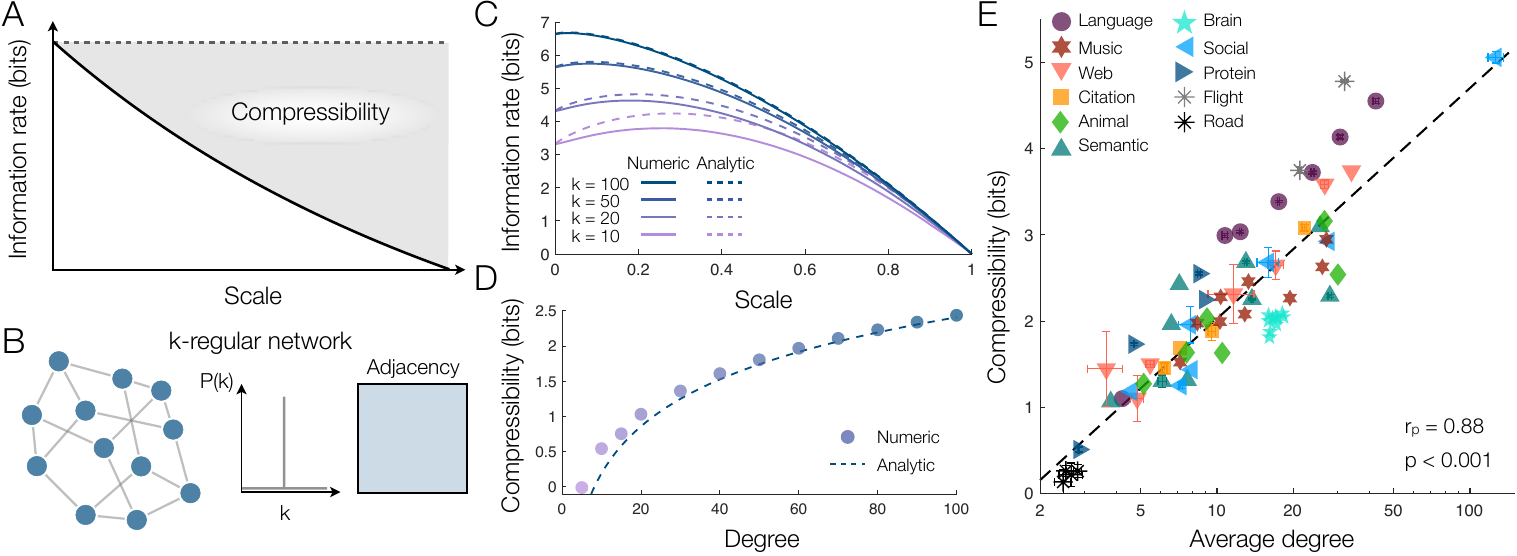}
\caption{\textbf{Quantifying compressibility.} (\textit{A}) The compressibility of a network (shaded region) is the area between the rate-distortion curve (solid line) and the entropy of random walks (dashed line). (\textit{B}) A $k$-regular network, characterized only by the requirement that all nodes have constant degree $k$. (\textit{C}) Rate-distortion curves $\bar{R}(S)$ for $k$-regular networks with different degrees $k$. (\textit{D}) Compressibility $\ubar{C}$ of $k$-regular networks versus degree $k$. In panels (\textit{C}) and (\textit{D}), solid lines and data points are averages over 50 randomly-generated networks, each of size $N = 10^3$, and dashed lines indicate analytic predictions (Eqs. \ref{Rk} and \ref{Ck}). (\textit{E}) Compressibility $\ubar{C}$ versus average degree for the real networks in Supporting Table 1. We note that average degree is plotted on a log scale. Dashed line indicates a logarithmic fit. For networks of size $N > 10^3$, data points and error bars represent means and standard deviations over 50 randomly-sampled subnetworks of $10^3$ nodes each (see Materials and Methods).}
\label{compressibility}
\end{figure*}

Intuitively, a network should be compressible if one can achieve a large reduction in the information rate at a given scale (Fig. \ref{rate_distortion}\textit{C}). However, rather than choosing a specific scale $S$ (equivalently, a specific number of clusters $n$), we would like our definition of compressibility to be a property of the network itself. We therefore define the compressibility of a network to be the amount of information that can be removed via compression, averaged across all scales,
\begin{equation}
\label{C}
C = H(\bm{x}) - \frac{1}{N}\sum_S R(S).
\end{equation}
Visually, the compressibility represents the area above a network's rate-distortion curve (Fig. \ref{compressibility}\textit{A}). In practice, plugging our tractable upper bound on the rate-distortion curve $\bar{R}(S)$ into Eq. \ref{C} yields a lower bound $\ubar{C}$, which (for simplicity) we will refer to as compressibility.

To make the notion of compressibility concrete, consider the class of random $k$-regular networks (Fig. \ref{compressibility}\textit{B}). On average, these networks have no structure (besides the requirement that nodes have uniform degree $k$), which allows us to derive an analytic approximation for the rate-distortion curve (see Supporting Information),
\begin{equation}
\label{Rk}
\bar{R}(S) \approx (1-S)^2\log k + S(1-S)\log N - S\log S.
\end{equation}
Each individual network, however, contains small structural variations, such as groups of nodes that are more tightly connected than expected. Generating random $k$-regular networks and computing their rate-distortion curves directly, we find that optimal compressions are able to capitalize on these structural variations (see Supporting Fig. 3), thereby achieving lower information rates than the approximation in Eq. \ref{Rk} (Fig. \ref{compressibility}\textit{C}). By contrast, as the degree $k$ increases, the networks become uniform in structure, and the analytic approximation becomes exact (Fig. \ref{compressibility}\textit{C}).

Using Eq. \ref{Rk}, one can predict the compressibility of $k$-regular networks. Specifically, noting that the entropy of $k$-regular networks is $\log k$ (see Materials and Methods), and approximating the average in Eq. \ref{C} by an integral over $S$, we arrive at the analytic form
\begin{equation}
\label{Ck}
\ubar{C} \approx \frac{2}{3}\log k - \frac{1}{6}\log N - \frac{1}{4\ln 2},
\end{equation}
which we verify numerically (Fig. \ref{compressibility}\textit{D}). We note that the compressibility grows logarithmically with the degree $k$, reflecting the fact that networks with larger degrees have more information to be removed via compression (see Materials and Methods). Indeed, computing the compressibility of the real networks in Supporting Table 1, we find precisely the same logarithmic dependence on the average degree (Fig. \ref{compressibility}\textit{E}). Furthermore, we verify that this logarithmic dependence generalizes to directed versions of the networks (Supporting Fig. 5) and is not simply due to our clustering heuristics (Supporting Fig. 6). These results demonstrate that the compressibility of a network increases predictably with average degree. But how does compressibility depend on the topology of a complex network?

\section*{Impact of network structure on compressibility}

\begin{figure*}[t!]
\centering
\includegraphics[width = \linewidth]{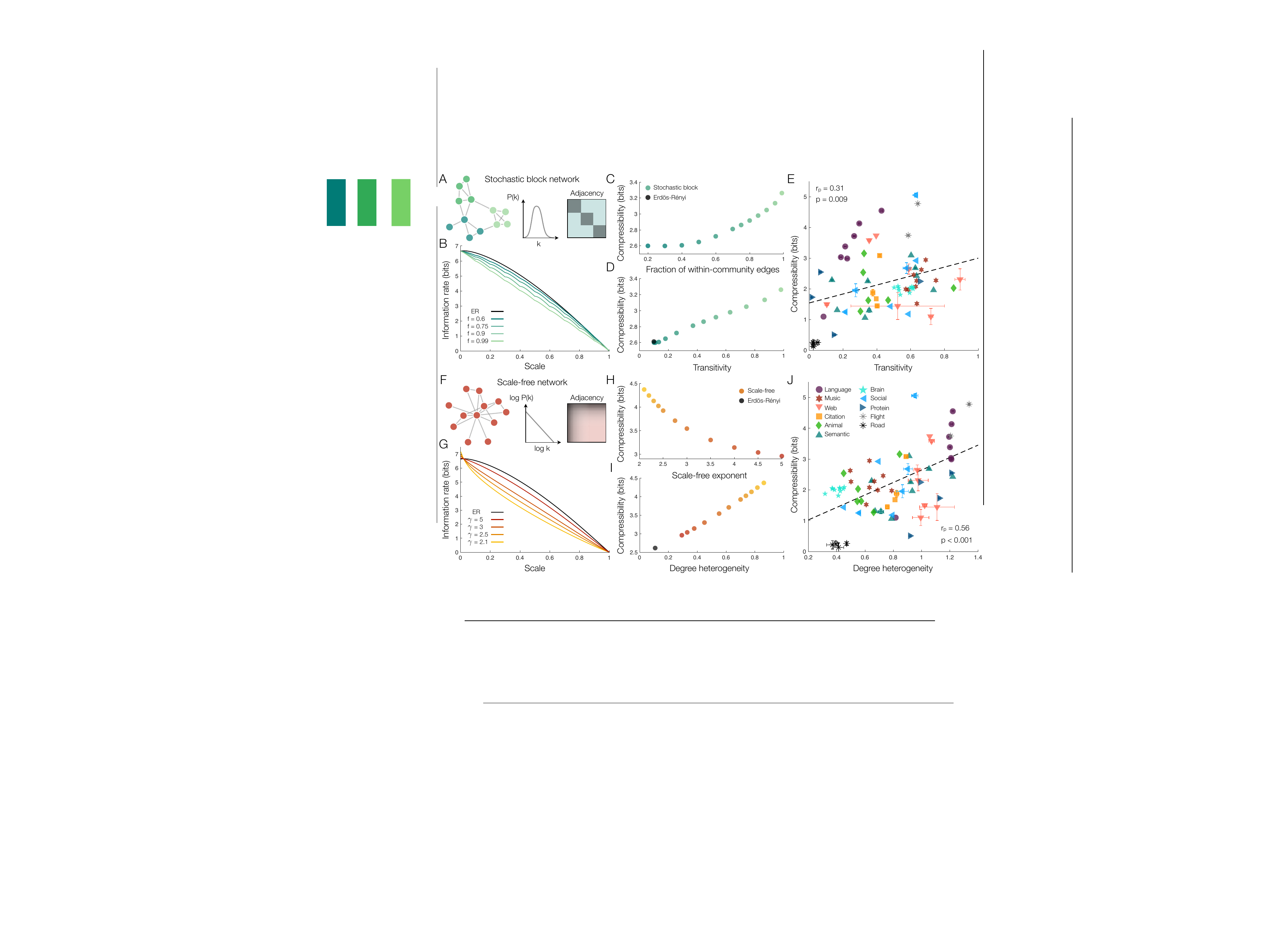}
\caption{\textbf{Compressibility increases with transitivity and degree heterogeneity.} (\textit{A}) Stochastic block network, characterized by dense connectivity within modules and sparse connectivity between modules. (\textit{B}) Rate-distortion curves $\bar{R}(S)$ for Erd\"{o}s-R\'{e}nyi networks (black line) and stochastic block networks (colored lines) with ten modules and different fractions $f$ of within-module edges. Undulations in the rate-distortion curves result from compressing each of the ten modules (see Supporting Fig. 3). (\textit{C}) Compressibility $\ubar{C}$ of stochastic block networks versus the fraction of within-module  edges $f$. (\textit{D}) Compressibility $\ubar{C}$ of stochastic block networks (colored points) and Erd\"{o}s-R\'{e}nyi networks (black point) versus transitivity (quantified by the average clustering coefficient). In panels (\textit{B}-\textit{D}), data reflect averages over 50 randomly-generated networks, each of size $N=10^3$ and average degree $\left<k\right> = 100$. (\textit{E}) Compressibility $\ubar{C}$ versus transitivity for the real networks in Supporting Table 1 with a linear best fit (dashed line). (\textit{F}) Scale-free network, characterized by a power-law degree distribution and the presence of high-degree hubs. (\textit{G}) Rate-distortion curves $\bar{R}(S)$ for Erd\"{o}s-R\'{e}nyi networks (black line) and scale-free networks (colored lines) with different scale-free exponents $\gamma$. (\textit{H}) Compressibility $\ubar{C}$ of scale-free networks versus the scale-free exponent $\gamma$. (\textit{I}) Compressibility $\ubar{C}$ of scale-free networks (colored points) and Erd\"{o}s-R\'{e}nyi networks (black point) versus degree heterogeneity $h$. In panels (\textit{G}-\textit{I}), data reflect averages over 50 networks generated using the static model \cite{Goh-01}, each of size $N=10^3$ and average degree $\left<k\right> = 100$. (\textit{J}) Compressibility $\ubar{C}$ versus degree heterogeneity for the real networks in Supporting Table 1 with a linear best fit (dashed line). In panels (\textit{E}) and (\textit{J}), for networks of size $N > 10^3$, data points and error bars represent means and standard deviations over 50 randomly-sampled subnetworks of $10^3$ nodes each (see Materials and Methods).}
\label{structure}
\end{figure*}

Based on the properties of optimal compressions (Fig. \ref{clustering}), we hypothesized that the compressibility of a network should increase with both (i) transitivity and (ii) degree heterogeneity. These two features are frequently observed across an array of real-world networks, from social, scientific, and biological interactions \cite{Ravasz-01, Tomassini-01, Ravasz-03} to the Internet \cite{Vazquez-02}, language \cite{Ravasz-01}, music \cite{Farbood-01}, and the brain \cite{Bassett-02}. Moreover, the combination of transitivity (with tightly-connected modules) and heterogeneous degrees (with well-connected hubs) defines hierarchical organization \cite{Ravasz-01}, which has been shown to support multi-scale representations of complex networks \cite{Sales-01, Rosvall-04} and enable efficient information processing in neural and communication systems \cite{Bassett-01, Lynn-07}. In fact, when encoding information about the world, the brain itself often employs hierarchical representations \cite{Balaguer-01, Diaconescu-01, Friston-04}. If correct, our hypotheses will lend to these perspectives a new outlook on the role of hierarchical structure: that it supports the efficient compression of complex networks.

To investigate the impact of transitivity and modular structure on compressibility, we consider a class of stochastic block networks (Fig. \ref{structure}\textit{A}), wherein nodes are grouped into modules of equal size and a specified fraction $f$ of the edges in the network connect nodes within the same module. We find that optimal compressions take advantage of this modular structure by clustering together nodes within the same module (see Supporting Fig. 3). Indeed, strengthening the modular structure -- that is, increasing the fraction $f$ of within-module edges -- decreases the rate-distortion curve $\bar{R}(S)$ (Fig. \ref{structure}\textit{B}). We therefore find that compressibility increases with both modularity (Fig. \ref{structure}\textit{C}) and transitivity (Fig. \ref{structure}\textit{D}). Importantly, these results on stochastic block networks generalize to real networks, with increases in transitivity yielding significant improvements in network compressibility (Fig. \ref{structure}\textit{E}).

To examine the dependence of compressibility on degree heterogeneity, we study scale-free networks (Fig. \ref{structure}\textit{F}), which have heavy-tailed degree distributions $P(k) \sim k^{-\gamma}$ characterized by a power-law exponent $\gamma$ \cite{Barabasi-01}. Optimal compressions exploit this heterogeneous structure by clustering together high-degree hub nodes (see Supporting Fig. 3). As $\gamma$ decreases, accentuating the heterogeneity in node degrees, the rate-distortion curve $\bar{R}(S)$ increases at small scales and decreases at intermediate and large scales (Fig. \ref{structure}\textit{G}). Both of these rate-distortion effects serve to improve the compressibility of scale-free networks (Fig. \ref{structure}\textit{H}). Moreover, rather than indirectly investigating the impact of heavy-tailed structure via the scale-free exponent $\gamma$, we can directly quantify the degree heterogeneity of a given network $h = \left<|k_i - k_j|\right>/\left<k\right>$, where $\left<|k_i - k_j|\right>$ is the absolute difference in degrees averaged over all pairs of nodes and $\left<k\right>$ is the average degree. We find that the compressibility of scale-free networks grows linearly with degree heterogeneity (Fig. \ref{structure}\textit{I}), a result that generalizes to real networks (Fig. \ref{structure}\textit{J}). Furthermore, we confirm that the dependencies of compressibility on both transitivity and degree heterogeneity extend to directed networks (Supporting Fig. 5) and are robust to our choice of clustering heuristics (Supporting Fig. 6).

Together, the above results demonstrate that transitivity and degree heterogeneity -- the two defining features of hierarchical organization -- increase the compressibility of complex networks. Indeed, in networks with explicit hierarchical organization (such as those examined in Ref. \cite{Ravasz-01}), we verify that optimal compressions capitalize on both modular structure and heterogeneous degrees in order to reduce the information rate (see Supporting Fig. 3). By contrast, for networks with uniform structure (such as Erd\"{o}s-R\'{e}nyi or $k$-regular networks), natural groupings of nodes do not exist (see Supporting Fig. 3), and therefore such networks are highly incompressible (Fig. \ref{structure}\textit{D,I}).

Interestingly, by focusing on specific families of networks, we discover variations in compressibility that reflect a network's specific function. Road networks, for example, exhibit the lowest transitivity and degree heterogeneity, and therefore the lowest compressibility, among the networks studied. This low compressibility is likely due to the fact that, unlike the other networks, road networks are confined to exist in two dimensions, severely constraining their topology \cite{Sperry-01}. Besides road networks, we find that protein interactions have the lowest transitivity and brain networks have the lowest degree heterogeneity, leading both classes of networks to be relatively incompressible. Interestingly, these two families are unique among the networks studied in that they are only encoded genetically and need not be represented cognitively by a human or animal. By contrast, language networks are highly compressible, perhaps reflecting the primary function of language as a means for encoding and communicating information. Thus, although many networks are encoded biologically, the pressure for these encodings to be efficient manifests to varying degrees in different families of networks, yielding a spectrum of compressibilities.

\section*{Discussion}

Complex networks perform an astonishing array of functions, which are supported by a multitude of topological structures. Many networks, however, are unified by a common constraint: that they rely on biological entities to encode them and pass them on. Encoding a network efficiently -- that is, striking an optimal balance between simplicity and accuracy -- requires compression, an insight that has provided information-theoretic perspectives on network structure \cite{Rosvall-02, Rosvall-03, Slonim-01}. Naturally, some networks should be more compressible than others, with structural regularities enabling efficient representations across multiple scales. To investigate this hypothesis, here we introduce a rate-distortion theory of network compression (Fig. \ref{rate_distortion}) and propose a quantitative definition for the compressibility of a network (Eq. \ref{C}; Fig. \ref{compressibility}\textit{A}). Applying our framework to a number of real and model networks, we demonstrate that network compressibility increases with both transitivity and degree heterogeneity (Fig. \ref{structure}), two features that together characterize hierarchical organization \cite{Ravasz-01}.

The interplay between network structure and compressibility hints at one possible factor contributing to the hierarchical organization observed in many real networks \cite{Ravasz-01, Sales-01, Ravasz-02, Dodds-02, Bassett-01, Lynn-07, Farbood-01, Vazquez-02, Bassett-02, Tomassini-01, Ravasz-03}: that it enables efficient representations across scales. But do the actual encodings themselves -- from representations in the minds of humans and animals to the information stored in the genetic code -- take advantage of this hierarchical organization? Moreover, do the encodings employed in nature approach the limit of optimal efficiency specified by rate-distortion theory? Answering these questions will require new exciting investigations into the information-theoretic forces that shape complex networks.

\matmethods{\subsection*{Entropy of random walks}

\hspace{0pt} Given a (possibly weighted, directed) network with adjacency matrix $G_{ij}$, the probability of one node $i$ transitioning to another node $j$ in a random walk is $P_{ij} = G_{ij}/k_i$, where $k_i = \sum_j G_{ij}$ is the (out) degree of node $i$ (Fig. \ref{rate_distortion}\textit{A}). The entropy of random walks is given by
\begin{equation}
\label{H}
H(\bm{x}) = -\sum_i \pi_i \sum_j P_{ij} \log P_{ij},
\end{equation}
where $\pi_i$ is the stationary distribution defined by the condition $\bm{\pi} = P^T \bm{\pi}$ (which we note is uniquely defined if the network is strongly-connected and aperiodic). For undirected networks, Eq. \ref{H} simplifies significantly. In this case, the stationary distribution is proportional to the node degrees $\pi_i = k_i/2E$, where $E = \frac{1}{2}\sum_{ij}G_{ij}$ is the number of edges in the network, and thus the entropy takes the form
\begin{equation}
H(\bm{x}) = \frac{1}{2E}\sum_i k_i \log k_i.
\end{equation}
If, in addition, the nodes have uniform degree $k$ (as in the $k$-regular networks in Fig. \ref{compressibility}) then the entropy equals $\log k$. For example, in the simple network in Fig. \ref{rate_distortion}, the nodes have uniform degree 4, and thus the entropy is 2 bits.

\subsection*{Bounding the information rate}

\hspace{0pt} After clustering a network, a random walk $\bm{x} = (x_1,x_2,\hdots)$ gives rise to a new sequence $\bm{y} = (y_1,y_2,\hdots)$, where $y_t$ is the cluster containing node $x_t$ (Fig. \ref{rate_distortion}\textit{B}). The information rate of this sequence is given by the mutual information $I(\bm{x},\bm{y})$, which for deterministic clusterings (such as those considered here) is equivalent to the entropy $H(\bm{y})$. However, even though the random walk $\bm{x}$ is Markovian (yielding a simple form for the entropy (Eq. \ref{H})), the clustered sequence $\bm{y}$ need not be \cite{Cover-01}, and thus it is generally difficult to derive an analytic form for $H(\bm{y})$.

Despite this hurdle, there exist simple bounds on the information rate $I(\bm{x},\bm{y}) = H(\bm{y})$, summarized by the inequalities
\begin{equation}
H(y_{t+1}\,|\,x_t) \le H(\bm{y}) \le H(y_{t+1}\,|\,y_t),
\end{equation}
where $H(y_{t+1}\,|\,x_t)$ and $H(y_{t+1}\,|\,y_t)$ are the conditional entropies of $y_{t+1}$ on $x_t$ and $y_t$, respectively \cite{Cover-01}. These bounds are tight at the minimum scale $S = 1/N$, when each cluster contains one node and so $H(\bm{y}) = H(\bm{x}) = H(x_{t+1}\,|\, x_t)$. The bounds are also tight at the maximum scale $S = 1$, when there is one cluster and so $H(\bm{y}) = H(y_{t+1}\,|\, x_t) = H(y_{t+1}\,|\, y_t) = 0$.

To compute the lower bound at intermediate scales, we begin with the conditional probability of node $i$ in the random walk $\bm{x}$ transitioning to cluster $c$ in the clustered sequence $\bm{y}$, $P_{ic} = \sum_{j\in c} P_{ij}$. Then, the lower bound is given by
\begin{equation}
\label{Ilow}
I(\bm{x},\bm{y}) \ge \ubar{I}(\bm{x},\bm{y}) = H(y_{t+1}\, |\, x_t) = - \sum_i \pi_i \sum_c P_{ic}\log P_{ic},
\end{equation}
where the second sum runs over all clusters $c$. Similarly, to compute the upper bound, we consider the probability of one cluster $c$ transitioning to another cluster $c'$,
\begin{equation}
P_{cc'} = \frac{1}{\pi_c}\sum_{i\in c} \pi_i \sum_{j\in c'} P_{ij},
\end{equation}
where $\pi_c = \sum_{i\in c} \pi_i$ is the stationary distribution over clusters. We then arrive at the following upper bound,
\begin{equation}
\label{Ihigh}
\hspace{-1pt} I(\bm{x},\bm{y}) \le \bar{I}(\bm{x},\bm{y}) = H(y_{t+1}\, |\,y_t) = -\sum_c \pi_c \sum_{c'} P_{cc'}\log P_{cc'},
\end{equation}
which is exact if the clustered sequence $\bm{y}$ is Markovian. In practice, when estimating the optimal information rate for a network, we minimize the upper bound in Eq. \ref{Ihigh} over clusterings, resulting in an upper bound $\bar{R}(S)$ on the rate-distortion curve.

The upper bound $\bar{I}(\bm{x},\bm{y})$ simplifies significantly for unweighted, undirected networks. In this case, the cluster transition probabilities take the form $P_{cc'} = G_{cc'}/k_c$, where $G_{cc'} = \sum_{i\in c}\sum_{j\in c'} G_{ij}$ is the induced network of clusters and $k_c = \sum_{i\in c}k_i$ is the sum of the degrees of the nodes in $c$. Recalling that the stationary distribution simplifies to $\pi_i = k_i/2E$, one can manipulate Eq. \ref{Ihigh} into the form
\begin{equation}
\label{Ihigh2}
\bar{I}(\bm{x},\bm{y}) = \frac{1}{2E}\bigg[ \sum_c k_c \log k_c - \sum_{cc'} G_{cc'}\log G_{cc'}\bigg].
\end{equation}
Under the further simplification of a clustering with one large cluster $c$ and $n-1$ minimal clusters of one node each (Fig. \ref{clustering}), this upper bound can be fashioned into Eq. \ref{Ibar}.

\subsection*{Clustering algorithm}

\hspace{0pt} To compute the rate-distortion curve $\bar{R}(S)$, we use an agglomerative clustering algorithm. Beginning with $n=N$ clusters (corresponding to the minimum scale $S = 1/N$), each containing an individual node, we iteratively combine pairs of clusters until we eventually arrive at one large cluster containing the entire network (corresponding to the maximum scale $S = 1$). At each step, we greedily select the pair of clusters to combine that minimizes the information rate $\bar{I}(\bm{x},\bm{y})$ (Eq. \ref{Ihigh}). However, rather than searching through all $n\choose 2$ pairs of clusters at each iteration (which would limit applications to small networks), we instead focus on a subset of $m$ pairs chosen through one of two heuristics.

The first heuristic, motivated by the observation that optimal clusterings tend to combine clusters with large degrees (Fig. \ref{clustering}\textit{E}), selects the $m$ pairs of clusters $c$ and $c'$ with the largest combined stationary probabilities $\pi_c + \pi_{c'}$. For unweighted, undirected networks, we note that this choice is equivalent to selecting the pairs of clusters with the largest combined degrees, since $\pi_c + \pi_{c'} = \frac{1}{2E}(k_c + k_{c'})$. The second heuristic, motivated by the fact that optimal compressions tend to form clusters with tight intra-cluster connectivity (Fig. \ref{clustering}\textit{D}), selects the pairs of clusters $c$ and $c'$ with the largest combined joint transition probabilities $\pi_c P_{cc'} + \pi_{c'}P_{c'c}$. For unweighted, undirected networks, we remark that this second heuristic is equivalent to selecting the pairs of clusters with the largest number of connecting edges, since $\pi_c P_{cc'} + \pi_{c'}P_{c'c} = \frac{1}{2E}(G_{cc'} + G_{c'c})$. In practice, we consider $m = 100$ pairs of clusters at each iteration. In Supporting Fig. 1, we compare these two heuristics to the brute-force approach that searches through all pairs of clusters at each iteration of the clustering algorithm. In addition to significantly speeding up the algorithm, we find that these two heuristics often yield more accurate estimates of the rate-distortion curve $R(S)$ than the brute-force implementation.

\subsection*{Network datasets}

\hspace{0pt} The networks analyzed in this paper are listed and described in Supporting Table 1. While we study unweighted, undirected versions of the networks in Figs. \ref{clustering}, \ref{compressibility}\textit{E}, \ref{structure}\textit{E}, and \ref{structure}\textit{J}, similar results hold for directed versions of the networks (Supporting Figs. 2 and 3). For networks of size $N \le 10^3$, we perform analyses directly. For larger networks with $N > 10^3$, we analyze 50 subnetworks of $10^3$ nodes each. Each subnetwork is generated by performing a random walk beginning at a randomly-selected node until $10^3$ nodes have been reached. This sampling method has been shown to give accurate estimates of network statistics \cite{Leskovec-03}.

\subsection*{Data and code availability} The data analyzed in this paper and the code used to perform the analyses are openly available at \texttt{github.com/ChrisWLynn/Network{\_}compressibility}.

}

\showmatmethods{} % Display the Materials and Methods section

\acknow{The authors thank Christopher Kroninger, Lia Papadopoulos, Dr. Pragya Srivastava, Mathieu Ouellet, and Dale Zhou for helpful feedback on earlier versions of this manuscript. C.W.L. acknowledges support from the James S. McDonnell Foundation 21$^{\text{st}}$ Century Science Initiative Understanding Dynamic and Multi-scale Systems - Postdoctoral Fellowship Award. The authors also acknowledge support from the John D. and Catherine T. MacArthur Foundation, the ISI Foundation, the Paul G. Allen Family Foundation, the Army Research Laboratory (W911NF-10-2-0022), the Army Research Office (Bassett-W911NF-14-1-0679, Falk-W911NF-18-1-0244, Grafton-W911NF-16-1-0474, DCIST- W911NF-17-2-0181), the Office of Naval Research, the National Institute of Mental Health (2-R01-DC-009209-11, R01-MH112847, R01-MH107235, R21-M MH-106799), and the National Science Foundation (NSF PHY-1554488, BCS-1631550, and NCS-FO-1926829).}

\showacknow{} % Display the acknowledgments section

\subsection*{CITATION DIVERSITY STATEMENT} 

\hspace{0pt} Recent work in several fields of science has identified a bias in citation practices such that papers from women and other minorities are under-cited relative to the number of such papers in the field \cite{Mitchell-01, Dion-01, Caplar-01, Maliniak-01, Dworkin-01, Bertolero-01}. Here we sought to proactively consider choosing references that reflect the diversity of the field in thought, form of contribution, gender, and other factors. We obtained predicted gender of the first and last author of each reference by using databases that store the probability of a name being carried by a woman \cite{Dworkin-01, Zhou-01}. By this measure (and excluding self-citations to the first and last authors of our current paper), our references contain $16\%$ woman(first)/woman(last), $17\%$ man/woman, $18\%$ woman/man, and $50\%$ man/man. This method is limited in that a) names, pronouns, and social media profiles used to construct the databases may not, in every case, be indicative of gender identity, and b) it cannot account for intersex, non-binary, or transgender people. Second, we obtained predicted racial/ethnic category of the first and last author of each reference by databases that store the probability of a first and last name being carried by an author of color \cite{Ambekar-01, Sood-01}. By this measure (and excluding self-citations), our references contain $9\%$ author of color(first)/author of color(last), $14\%$ white author/author of color, $15\%$ author of color/white author, and $62\%$ white author/white author. This method is limited in that a) names, Census entries, and Wikipedia profiles used to make the predictions may not be indicative of racial/ethnic identity, and b) it cannot account for Indigenous and mixed-race authors, or those who may face differential biases due to the ambiguous racialization or ethnicization of their names. We look forward to future work that could help us to better understand how to support equitable practices in science.

\newpage
\clearpage

\section*{Supporting Information}

\section{Introduction}

In this Supporting Information, we provide extended analysis and discussion to support the results presented in the main text. In Sec. \ref{edges}, we discuss the information costs of the different edge types in Fig. 2\textit{C} in the main text. In Sec. \ref{kreg}, we derive an analytic approximation of the rate-distortion curve for $k$-regular networks (Eq. 3 and Fig. 3\textit{C} in the main text). In Sec. \ref{speedup}, we discuss the different heuristics used to speed up the clustering algorithm and compare their estimates of optimal information rates with those of a brute-force implementation. In Sec. \ref{bounds}, we demonstrate that the tractable upper bound on the rate-distortion curve $\bar{R}(S)$ used throughout the main text provides a reasonable approximation to the true rate-distortion curve. In Sec. \ref{rand_nets}, we study the structure of optimal compressions for the model networks analyzed in the main text. In Sec. \ref{directed}, we demonstrate that the central results from the main text generalize to directed networks. In Sec. \ref{robust}, we show that the dependencies of compressibility on average degree, transitivity, and degree heterogeneity do not depend on the heuristics used to speed up the clustering algorithm. Finally, in Sec. \ref{datasets}, we list the real networks analyzed in this work and describe how we sample large networks.

\section{Information content of different edges}
\label{edges}

When analyzing optimal clusterings (that is, clusterings that minimize the information rate $\bar{I}(\bm{x},\bm{y})$, Eq. 9 in the main text), we find that they tend to consist of one large cluster containing $N - n + 1 = SN$ nodes and $n-1$ clusters containing one node each (see Fig. 2\textit{A}, \textit{B} in the main text). This observation allows us to group the edges in a network into three categories (Fig. 2\textit{C} in the main text): those connecting nodes \textit{within} $c$, those connecting nodes \textit{outside} of $c$, and those on the \textit{boundary} of $c$ (connecting nodes within $c$ to nodes outside of $c$).

In order to predict the structure of the one large cluster $c$, we wish to compare the contributions of different edge types to the information rate $\bar{I}(\bm{x},\bm{y})$. For an unweighted, undirected network with adjacency matrix $G_{ij}$, and a clustering with one large cluster $c$, the information rate can be written as
\begin{align}
\label{I}
\bar{I}(\bm{x},\bm{y}) &= \frac{1}{2E}\left[ \sum_{i\not\in c} k_i \log k_i + k_c \log k_c \right. \nonumber \\
&\quad\quad\quad\quad \left. - 2\sum_{i\not\in c} G_{ic}\log G_{ic} - G_{cc} \log G_{cc}\right],
\end{align}
where the sums run over all nodes $i$ not in $c$, $E = \frac{1}{2}\sum_{ij} G_{ij}$ is the number of edges in the network, $k_i = \sum_j G_{ij}$ is the degree of node $i$, $k_c = \sum_{i\in c} k_i$ is the combined degrees of nodes in $c$, $G_{ic} = \sum_{j\in c} G_{ij}$ is the number of edges connecting a node $i$ to nodes in $c$, and $G_{cc} = \sum_{ij\in c} G_{ij}$ is the number of edges connecting nodes within $c$ (see Materials and Methods in the main text).

\subsection*{Within versus boundary edges}

If we add an edge between two nodes within $c$, then both $k_c$ and $G_{cc}$ increase by two, and the change in the information rate is given by
\begin{align}
\label{within}
\Delta \bar{I}^{\text{within}} &= \frac{1}{2E}\big[(k_c + 2)\log(k_c + 2) - (G_{cc}+2)\log(G_{cc}+2) \nonumber \\
&\quad\quad\quad\quad - k_c\log k_c + G_{cc}\log G_{cc}\big] \nonumber \\
&\approx \frac{1}{2E}\left(2\log k_c - 2\log G_{cc}\right),
\end{align}
where the approximation follows by letting $\log (k_c + 2) \approx \log k_c$ and $\log (G_{cc} + 2) \approx \log G_{cc}$. By contrast, adding an edge on the boundary of $c$ (say, connecting a node $i$ outside of $c$ to a node in $c$) yields a contribution to the information rate of
\begin{align}
\label{boundary}
\Delta \bar{I}^{\text{boundary}} &= \frac{1}{2E}\big[(k_i + 1)\log(k_i + 1) + (k_c+ 1)\log(k_c + 1) \nonumber \\
&\quad\quad\quad\quad - 2(G_{ic} + 1)\log (G_{ic}+1) - k_i\log k_i \nonumber \\
&\quad\quad\quad\quad - k_c\log k_c + 2G_{ic}\log G_{ic} \big] \nonumber \\
&\approx \frac{1}{2E}\left(\log k_i + \log k_c - 2\log G_{ic}\right),
\end{align}
where the approximation follows from $\log (k_i + 1) \approx \log k_i$, $\log (k_c +1) \approx \log k_c$, and $\log (G_{ic} + 1) \approx \log G_{ic}$.

It is clear that $\Delta \bar{I}^{\text{within}}$ will be less than $\Delta \bar{I}^{\text{boundary}}$ if $k_c /G_{cc}^2 \le k_i/G_{ic}^2$. Given a random selection of nodes to include in $c$, both the fraction $G_{cc}/k_c$ of edges emanating from $c$ that end in $c$ and the fraction $G_{ic}/k_i$ of edges emanating from $i$ that end in $c$ are approximated by the proportional size of $c$; namely, $SN/N = S$. Because $G_{cc} \gg G_{ic}$, for a randomly-selected cluster $c$ we have $k_c /G_{cc}^2 \approx 1/SG_{cc} \ll 1/S G_{ic} \approx k_i/G_{ic}^2$. Thus, the information cost of edges within $c$ ($\Delta\bar{I}^{\text{within}}$) is likely to be lower than the cost of an edge on the boundary of $c$ ($\Delta\bar{I}^{\text{boundary}}$), leading to the prediction that optimal clusters $c$ will seek to include tightly-knit communities with sparse connectivity to the rest of the network. We confirm this prediction in real networks (Fig. 2\textit{D} in the main text), and we further demonstrate that modular structure and tight clustering serve to increase network compressibility (Fig. 4\textit{A}-\textit{E} in the main text).

\subsection*{Within versus outside edges}

Now consider an edge connecting two nodes $i$ and $j$ outside of $c$. Adding such an edge increases the information rate by an amount
\begin{align}
\label{outside}
\Delta \bar{I}^{\text{outside}} &= \frac{1}{2E}\big[(k_i + 1)\log(k_i + 1) + (k_j+ 1)\log(k_j + 1) \nonumber \\
&\quad\quad\quad\quad - k_i\log k_i - k_j\log k_j\big] \nonumber \\
&\approx \frac{1}{2E}\left(\log k_i + \log k_j\right),
\end{align}
where the approximation follows from $\log (k_i + 1) \approx \log k_i$ and $\log (k_j +1) \approx \log k_j$. Comparing Eqs. \ref{within} and \ref{outside}, we see that $\Delta \bar{I}^{\text{within}}$ will be less than $\Delta \bar{I}^{\text{outside}}$ if $k_c^2 /G_{cc}^2 \le k_i k_j$. Based on the above result that optimal clusters $c$ tend to include tight within-cluster connectivity, we know that the proportion $G_{cc}/k_c$ is greater than the scale $S$ (see Fig. 2\textit{D} in the main text). Thus, we will have $k_c^2 /G_{cc}^2 \le k_i k_j$ (and therefore $\Delta \bar{I}^{\text{within}} \le \Delta \bar{I}^{\text{outside}}$) across most scales (specifically, for all scales $S \ge 1/\sqrt{k_i k_j}$). This result indicates that optimal clusters $c$ will seek to include high-degree nodes and exclude low-degree nodes, which we confirm in Fig. 2\textit{E} in the main text. Moreover, because compression leverages differences in node degrees, we find that networks with heterogeneous (or heavy-tailed) degrees are highly compressible (Fig. 4\textit{F}-\textit{J} in the main text).

\section{Rate-distortion curve for $k$-regular networks}
\label{kreg}

We wish to derive an analytic approximation to the rate-distortion curve $\bar{R}(S)$ for $k$-regular networks (Fig. 3\textit{B} in the main text). For unweighted, undirected networks and a clustering including one large cluster $c$ of size $SN$, the information rate $\bar{I}(\bm{x},\bm{y})$ is given in Eq. \ref{I}. Let us examine each term individually:
\begin{itemize}
\item The large cluster $c$ contains $N-n+1 = SN$ nodes, and so the number of nodes outside of $c$ is $n-1 = (1-S)N$. Since each node in the network has degree $k$, we have
\begin{equation}
\label{reg1}
\sum_{i\not\in c} k_i \log k_i = (1-S)Nk\log k,
\end{equation}
and
\begin{equation}
\label{reg2}
k_c\log k_c = SNk\log (SNk).
\end{equation}

\item Assuming that each edge has a probability $S$ of connecting to nodes in $c$, we can approximate $G_{ic} \approx Sk_i = Sk$ and $G_{cc} \approx Sk_c = S^2Nk$. Thus, we can approximate
\begin{equation}
\label{reg3}
\sum_{i\not\in c} G_{ic}\log G_{ic} \approx (1-S)NSk \log (Sk),
\end{equation}
and
\begin{equation}
\label{reg4}
G_{cc}\log G_{cc} \approx S^2Nk \log (S^2Nk).
\end{equation}
\end{itemize}

\noindent Plugging Eqs. \ref{reg1}-\ref{reg4} into Eq. \ref{I}, and noting that $2E = kN$, after some algebra we arrive at an analytic approximation to the rate-distortion curve for a $k$-regular network,
\begin{equation}
\bar{R}(S) \approx (1-S)^2\log k + S(1-S)\log N - S\log S.
\end{equation}
We demonstrate that this prediction provides a good approximation to true rate-distortion curves, and becomes accurate in the high-degree limit (Fig. 3\textit{C} in the main text).

\section{Speeding up the clustering algorithm}
\label{speedup}

To compute the rate-distortion curve $\bar{R}(S)$ for a given network, one must identify clusterings that minimize the information rate $\bar{I}(\bm{x},\bm{y})$ across all scales $S$ (equivalently, for all numbers of clusters $n = N,\hdots,1$). Here, we employ a greedy clustering algorithm that iteratively combines pairs of clusters so as to minimize the information rate (see Materials and Methods in the main text). Specifically, beginning with $n = N$ clusters (one for each node in the network), we attempt to combine different pairs of clusters and compute the resulting changes to the information rate $\bar{I}(\bm{x},\bm{y})$. Combining the pair of clusters that yields the largest reduction in information rate, we arrive at a new clustering with $n-1$ clusters. Repeating this process for all numbers of clusters $n = N,\hdots, 1$, we arrive at estimates of the optimal clusterings and information rates across all scales $S$.

A brute force implementation of this algorithm would attempt to combine all ${n \choose 2} = O(n^2) = O(N^2)$ pairs of clusters at each iteration. For each pair of clusters, one would then compute the new information rate
\begin{equation}
\label{Ibar_supp}
\bar{I}(\bm{x},\bm{y}) = -\sum_c \pi_c \sum_{c'} P_{cc'} \log P_{cc'},
\end{equation}
where $\pi_c = \sum_{i\in c} \pi_i$ is the stationary distribution over clusters, and 
\begin{equation}
P_{cc'} = \frac{1}{\pi_c} \sum_{i\in c} \pi_i \sum_{j\in c'} P_{ij}
\end{equation}
is the conditional probability of transitioning from cluster $c$ to cluster $c'$. Since Eq. \ref{Ibar_supp} involves summing over all pairs of clusters $c$ and $c'$, computing the information rate requires $O(n^2) = O(N^2)$ computations. Finally, the algorithm repeats this process for all numbers of clusters $n = N,\hdots,1$, requiring a total of $O(N^5)$ computations. This $N^5$ dependence limits a na\"{i}ve implementation of our clustering algorithm to small networks. In what follows, we will show how to reduce this size dependence to $N^2$, significantly improving the efficiency of the algorithm and enabling applications to networks of reasonable size.

\subsection*{Change in information rate}

As discussed above, a na\"{i}ve implementation of the algorithm would re-compute the information rate (Eq. \ref{Ibar_supp}) after attempting to combine each pair of clusters, requiring $O(n^2)$ computations. However, we only require the \textit{change} in the information rate, a computation that we will see takes $O(n)$ time.

Consider attempting to combine two clusters $\alpha$ and $\beta$. Before combining the clusters, the information rate is given by
\begin{align}
\bar{I}^{\text{old}} = &-\sum_{c\neq \alpha,\beta} \pi_c \sum_{c' \neq \alpha, \beta} P_{cc'}\log P_{cc'} \nonumber \\
&- \sum_{c\neq \alpha,\beta} \pi_c (P_{c\alpha}\log P_{c\alpha} + P_{c\beta}\log P_{c\beta}) \nonumber \\
& - \pi_\alpha \sum_{c\neq \alpha,\beta}P_{\alpha c}\log P_{\alpha c} - \pi_\beta \sum_{c\neq \alpha,\beta}P_{\beta c}\log P_{\beta c} \nonumber \\
&- \pi_\alpha (P_{\alpha\alpha}\log P_{\alpha\alpha} + P_{\alpha\beta}\log P_{\alpha\beta}) \nonumber \\
&- \pi_\beta(P_{\beta\beta}\log P_{\beta\beta} + P_{\beta\alpha}\log P_{\beta\alpha}).
\end{align}
After combining the clusters $\alpha$ and $\beta$, the new information rate is given by
\begin{align}
\bar{I}^{\text{new}} &= -\sum_{c\neq \alpha,\beta} \pi_c \sum_{c' \neq \alpha, \beta} P_{cc'}\log P_{cc'} \nonumber \\
&\quad - \sum_{c\neq \alpha,\beta} \pi_c (P_{c\alpha} + P_{c\beta}) \log (P_{c\alpha} + P_{c\beta}) \nonumber \\
&\quad -(\pi_\alpha + \pi_\beta) \sum_{c\neq \alpha, \beta} \frac{\pi_\alpha P_{\alpha c} + \pi_\beta P_{\beta c}}{\pi_\alpha + \pi_\beta} \log \frac{\pi_\alpha P_{\alpha c} + \pi_\beta P_{\beta c}}{\pi_\alpha + \pi_\beta} \nonumber \\
&\quad - (\pi_\alpha + \pi_\beta)\frac{\pi_\alpha (P_{\alpha \alpha} + P_{\alpha\beta}) + \pi_\beta (P_{\beta\beta} + P_{\beta\alpha})}{\pi_\alpha + \pi_\beta} \nonumber \\
&\quad\quad\quad \cdot \log \frac{\pi_\alpha (P_{\alpha \alpha} + P_{\alpha\beta}) + \pi_\beta (P_{\beta\beta} + P_{\beta\alpha})}{\pi_\alpha + \pi_\beta} \nonumber \\
&= -\sum_{c\neq \alpha,\beta} \pi_c \sum_{c' \neq \alpha, \beta} P_{cc'}\log P_{cc'} \nonumber \\
&\quad - \sum_{c\neq \alpha,\beta} \pi_c (P_{c\alpha} + P_{c\beta}) \log (P_{c\alpha} + P_{c\beta}) \nonumber \\
&\quad - \sum_{c\neq \alpha, \beta} (\pi_\alpha P_{\alpha c} + \pi_\beta P_{\beta c}) \log \frac{\pi_\alpha P_{\alpha c} + \pi_\beta P_{\beta c}}{\pi_\alpha + \pi_\beta} \nonumber \\
&\quad - (\pi_\alpha (P_{\alpha \alpha} + P_{\alpha\beta}) + \pi_\beta (P_{\beta\beta} + P_{\beta\alpha})) \nonumber \\
& \quad\quad\quad \cdot \log \frac{\pi_\alpha (P_{\alpha \alpha} + P_{\alpha\beta}) + \pi_\beta (P_{\beta\beta} + P_{\beta\alpha})}{\pi_\alpha + \pi_\beta}.
\end{align}
Thus, the change in the information after combining clusters $\alpha$ and $\beta$ is given by
\begin{align}
\label{dIbar}
\Delta \bar{I} &= \bar{I}^{\text{new}} - \bar{I}^{\text{old}} \nonumber \\
&= - \sum_{c\neq \alpha,\beta} \pi_c (P_{c\alpha} + P_{c\beta}) \log (P_{c\alpha} + P_{c\beta}) \nonumber \\
&\quad - \sum_{c\neq \alpha, \beta} (\pi_\alpha P_{\alpha c} + \pi_\beta P_{\beta c}) \log \frac{\pi_\alpha P_{\alpha c} + \pi_\beta P_{\beta c}}{\pi_\alpha + \pi_\beta} \nonumber \\
&\quad - (\pi_\alpha (P_{\alpha \alpha} + P_{\alpha\beta}) + \pi_\beta (P_{\beta\beta} + P_{\beta\alpha})) \nonumber \\
& \quad\quad\quad \cdot \log \frac{\pi_\alpha (P_{\alpha \alpha} + P_{\alpha\beta}) + \pi_\beta (P_{\beta\beta} + P_{\beta\alpha})}{\pi_\alpha + \pi_\beta} \nonumber \\
&\quad + \sum_{c\neq \alpha,\beta} \pi_c (P_{c\alpha}\log P_{c\alpha} + P_{c\beta}\log P_{c\beta}) \nonumber \\
& \quad + \pi_\alpha \sum_{c\neq \alpha,\beta}P_{\alpha c}\log P_{\alpha c} + \pi_\beta \sum_{c\neq \alpha,\beta}P_{\beta c}\log P_{\beta c} \nonumber \\
& \quad + \pi_\alpha (P_{\alpha\alpha}\log P_{\alpha\alpha} + P_{\alpha\beta}\log P_{\alpha\beta}) \nonumber \\
& \quad - \pi_\beta(P_{\beta\beta}\log P_{\beta\beta} + P_{\beta\alpha}\log P_{\beta\alpha}).
\end{align}
Although Eq. \ref{dIbar} appears more complicated than Eq. \ref{Ibar_supp}, we remark that it only requires summing over the different clusters $c$ rather than all pairs of clusters $c$ and $c'$. Therefore, by computing the change in information rate rather than the information rate itself, we reduce the number of computations from $O(n^2)$ to $O(n)$.

\subsection*{Heuristics for cluster selection}

In the na\"{i}ve implementation of the clustering algorithm, one searches through all ${n \choose 2} = O(n^2)$ pairs of clusters at each iteration to find the pair whose combination yields the largest decrease in the information rate. Here, we instead propose two heuristics for selecting a subset of $m$ pairs of clusters at each iteration, thereby reducing the number of pairs from $O(n^2)$ to $m$. For all results here and in the main text, we consider $m = 100$ pairs of clusters at each iteration of the clustering algorithm.

The first heuristic is motivated by the observation that optimal clusterings tend to include one large cluster with high-degree nodes (Fig. 2\text{E} in the main text). In undirected networks, the stationary distribution over clusters is proportional to the cluster degrees, such that $\pi_c = k_c/2E$, where $k_c = \sum_{i\in c}k_i$ is the degree of cluster $c$ and $E = \frac{1}{2}\sum_{ij} G_{ij}$ is the number of edges in the network. We therefore select the $m$ pairs of clusters $c$ and $c'$ with the largest combined stationary probabilities $\pi_c + \pi_{c'}$; for undirected networks, this process is equivalent to selecting the pairs of clusters with the largest combined degrees $k_c + k_{c'}$.

The second heuristic is motivated by the observation that the one large cluster tends to include nodes that are tightly connected to one another (Fig. 2\textit{D} in the main text). In undirected networks, the joint transition probability from one cluster $c$ to another $c'$ is proportional to the number of edges between the clusters, such that $\pi_c P_{cc'} = G_{cc'}/2E$, where $G_{cc'} = \sum_{i\in c, j\in c'} G_{ij}$ is the number of edges connecting nodes in $c$ to nodes in $c'$. We therefore select the $m$ pairs of clusters $c$ and $c'$ with the largest joint transition probabilities $\pi_c P_{cc'} + \pi_{c'}P_{c'c}$; for undirected networks, this process is equivalent to selecting the pairs of clusters with the largest number of connecting edges $G_{cc'} + G_{c'c}$.

To evaluate the performance of these two heuristics, we compare the rate-distortion curves for the real networks in Supporting Table 1 computed using (i) the brute-force approach that attempts to combine all pairs of clusters, (ii) the stationary distribution heuristic, and (iii) the joint transition probability heuristic. Since each algorithm computes an upper bound on the rate-distortion curve $\bar{R}(S)$ for a given network, whichever algorithm returns a smaller upper bound will have achieved a more accurate estimate of the true rate-distortion curve $R(S)$. Consider, for example, the social network of bottlenose dolphins \cite{Lusseau-01}. For scales $S \lesssim 0.8$, we find that the stationary distribution heuristic provides a lower (and thus more accurate) upper bound on the rate-distortion curve $\bar{R}(S)$ than both the brute-force algorithm and the joint transition probability heuristic (Supporting Fig. \ref{heuristics}\textit{A}). By contrast, for scales $S \gtrsim 0.8$, the joint transition probability heuristic provides the most accurate estimate of the rate-distortion curve (Supporting Fig. \ref{heuristics}\textit{A}). Notably, even though both heuristics only consider a limited set of cluster pairs, they both produce rate-distortion estimates that are comparable in accuracy to, if not more accurate than, the brute-force approach that searches through all pairs at each iteration.

\setcounter{figure}{0}
\begin{figure*}[t]
\centering
\includegraphics[width = \textwidth]{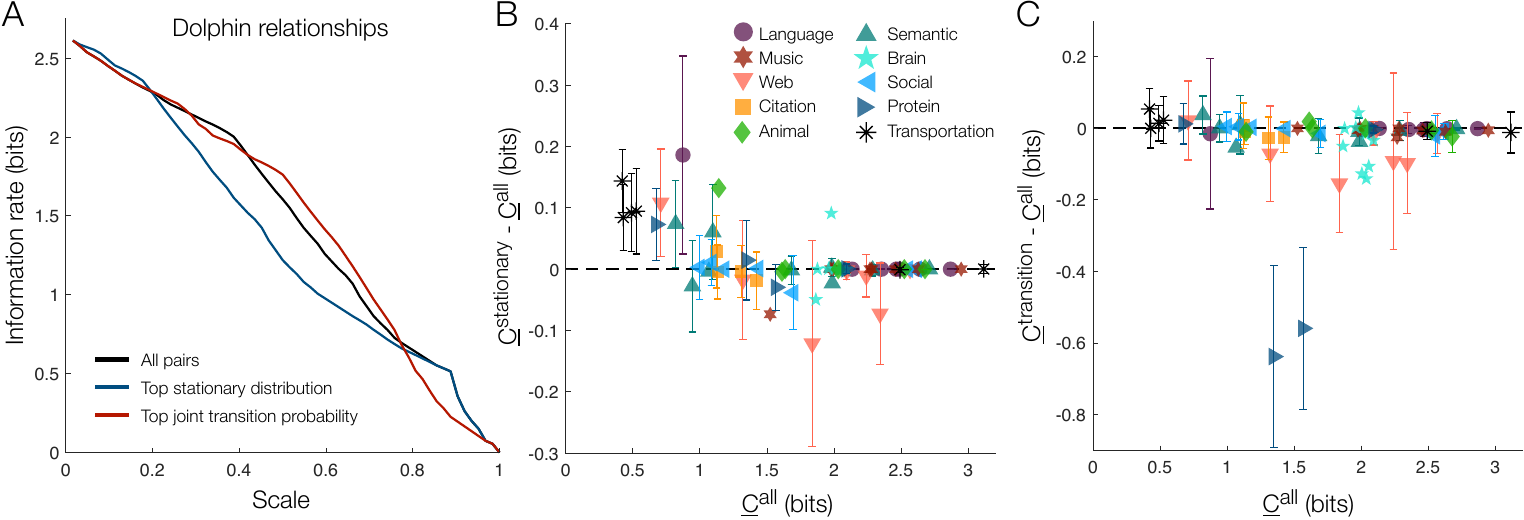}
\captionsetup{labelformat=empty}
\caption{\textbf{Supporting Fig. \ref{heuristics}. Performance of clustering heuristics.} (\textit{A}) Rate-distortion curves for a network of interactions between bottlenose dolphins \cite{Lusseau-01} computed using the clustering algorithm with the brute-force strategy that tests all pairs of clusters at each iteration (black), the top stationary distribution heuristic (blue), and the top joint transition probability heuristic (red). (\textit{B}) The difference between compressibility computed using the stationary distribution heuristic $C^{\text{stationary}}$ and that computed using all pairs of clusters $C^{\text{all}}$ as a function of the compressibility $C^{\text{all}}$ for the real networks in Supporting Table 1. (\textit{C}) Difference between compressibility computed using the joint transition probability heuristic $C^{\text{transition}}$ and that computed using all pairs of clusters $C^{\text{all}}$ as a function of the compressibility $C^{\text{all}}$. In panels (\textit{B}) and (\textit{C}), for networks of size $N > 100$, data points and error bars represent averages and standard deviations, respectively, over 50 subnetworks of 100 nodes each sampled using random walks beginning at random seed nodes (see Materials and Methods in the main text). \label{heuristics}}
\end{figure*}

To compare the accuracy of the different algorithms (that is, to determine which algorithm provides a lower upper bound $\bar{R}(S)$ across all scales) we can compare their compressibility estimates,
\begin{equation}
\label{C_supp}
\ubar{C} = H(\bm{x}) - \frac{1}{N}\sum_S \bar{R}(S).
\end{equation}
If one algorithm produces a lower (that is, more accurate) rate-distortion estimate $\bar{R}(S)$ on average across all scales $S$, then one will arrive at a larger lower bound on the compressibility $\ubar{C}$. In Supporting Fig. \ref{heuristics}\textit{B}, we see that the compressibility computed using the stationary distribution heuristic $\ubar{C}^{\text{stationary}}$ is almost always as large as (if not larger than) that computed using all pairs of clusters $\ubar{C}^{\text{all}}$ for the real networks in Supporting Table 1. Similarly, in Supporting Fig. \ref{heuristics}\textit{C} we see the joint transition probability heuristic provides compressibility estimates $\ubar{C}^{\text{transition}}$ that are nearly as accurate as $\ubar{C}^{\text{all}}$ across most of the real networks. In fact, the brute-force compressibility estimates $\ubar{C}^{\text{all}}$ are significantly larger (that is, more accurate) than both of the heuristic estimates $\ubar{C}^{\text{stationary}}$ and $\ubar{C}^{\text{transition}}$ for only 10 of the 69 real networks.

Together, these results demonstrate that the two cluster-selection heuristics provide rate-distortion estimates that are comparable, if not more accurate, than the brute-force clustering algorithm. Moreover, these heuristics reduce the search for an optimal pair of clusters at each iteration from $O(n^2) = O(N^2)$ pairs to $m$ pairs. In combination with the speed-up in Sec. \ref{speedup}, this reduces the total run-time of the clustering algorithm from $O(N^5)$ to $O(mN^2)$, thereby allowing applications to networks of reasonable size.

\section{Errors in information bounds}
\label{bounds}

In the main text, rather than computing the information rate $I(\bm{x},\bm{y})$ of a given clustering directly, we instead consider a tractable upper bound $\bar{I}(\bm{x},\bm{y})$ (see Materials and Methods in the main text). For a given network, we minimize $\bar{I}(\bm{x},\bm{y})$ across all scales to arrive at an upper bound on the rate-distortion curve $\bar{R}(S)$, which we then use to compute a lower bound $\ubar{C}$ on the compressibility (Eq. \ref{C}). The upper bound $\bar{R}(S)$ on the rate-distortion curve and the corresponding lower bound $\ubar{C}$ on the compressibility together form the basis of our main results (Figs. 3 and 4 in the main text). Therefore, it is important to verify that these bounds provide reasonable approximations to the true rate-distortion curves $R(S)$ and compressibilities $C$ of the networks analyzed in the main text.

To investigate the accuracy of the upper bound on the rate-distortion curve $\bar{R}(S)$, we consider a tractable lower bound $\ubar{R}(S)$ (see Materials and Methods in the main text). Importantly, the true rate-distortion curve $R(S)$ lies between the upper and lower bounds, such that $\bar{R}(S) \ge R(S) \ge \ubar{R}(S)$. Thus, the error in the upper bound ($\bar{R}(S) - R(S)$) is no larger than the difference between the upper and lower bounds ($\bar{R}(S) - \ubar{R}(S)$). Put simply, if the difference $\bar{R}(S) - \ubar{R}(S)$ between the bounds is small, then we know that the error $\bar{R}(S) - R(S)$ is even smaller.

\begin{figure*}[t]
\centering
\includegraphics[width = \textwidth]{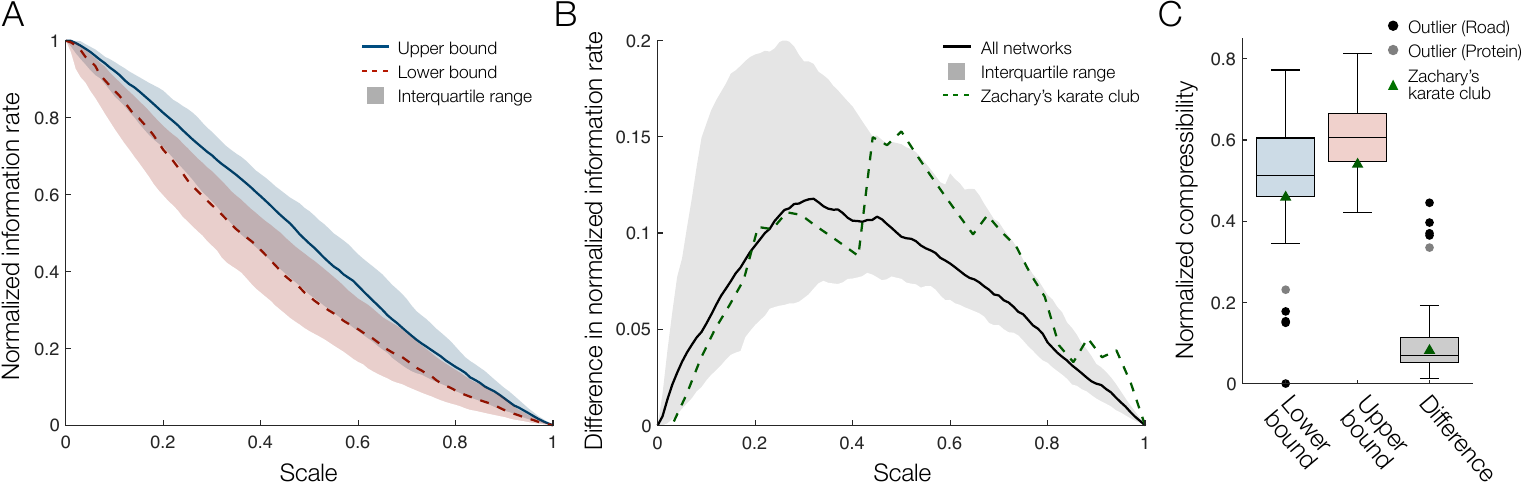}
\captionsetup{labelformat=empty}
\caption{\textbf{Supporting Fig. \ref{error}. Errors in rate-distortion curves and compressibilities.} (\textit{A}) Upper bounds $\bar{R}(S)$ (blue) and lower bounds $\ubar{R}(S)$ (red) on the rate-distortion curves of the real networks in Supporting Table 1, normalized by the entropy $H(\bm{x})$ of each network. Lines represent medians over the real networks, and shaded regions reflect interquartile ranges. (\textit{B}) Difference $\bar{R}(S) - \ubar{R}(S)$ between the upper and lower bounds on the rate-distortion curves, normalized by the entropy $H(\bm{x})$. The black line and shaded region represent the median and interquartile range over the real networks in Supporting Table 1, respectively, and the green dashed line indicates Zachary's karate club (see Fig. 2\textit{A} in the main text). (\textit{C}) Box plots (over the real networks in Supporting Table 1) of the lower bound on the compressibility $\ubar{C}$ (computed using the upper bound on the rate-distortion curve $\bar{R}(S)$; blue), the upper bound on the compressibility $\bar{C}$ (computed using $\ubar{R}(S)$; red), and the difference $\bar{C} - \ubar{C}$ (grey), all normalized by the entropy $H(\bm{x})$. Whiskers have a maximum length of 1.5 times the interquartile range. Outliers are indicated by circles, either black (for road networks) or grey (for protein networks). Green triangles illustrate compressibility values for Zachary's karate club.\label{error}}
\end{figure*}

For all networks, one can show that the upper and lower bounds are equal (and therefore exact) at both the minimum scale $S = 1/N$ and the maximum scale $S=1$ (see Materials and Methods in the main text). Moreover, in the main text, we demonstrated for Zachary's karate club \cite{Zachary-01} that the upper bound $\bar{R}(S)$ remains close to the lower bound $\ubar{R}(S)$ across all intermediate scales (Fig. 2\textit{A} in the main text). In order to compare the rate-distortion bounds across different networks, here we normalize $\bar{R}(S)$ and $\ubar{R}(S)$ by the entropy $H(\bm{x})$ of each network. Notably, we find that a tight correspondence between the upper and lower bounds is not unique to Zachary's karate club, but instead is a general feature of the real networks in Supporting Table 1 (Supporting Fig. \ref{error}\textit{A}). Indeed, the maximum difference $\bar{R}(S) - \ubar{R}(S)$ between the upper and lower bounds tends to only reach about 10\% of the entropy of a network (Supporting Fig. \ref{error}\textit{B}). For comparison, the difference between bounds is actually larger in Zachary's karate club (reaching 15\% of the entropy) than the typical network in Supporting Table 1 (Supporting Fig. \ref{error}\textit{B}). These results establish that the upper bound $\bar{R}(S)$ provides a good approximation to the true rate-distortion curve $R(S)$.

We now investigate the accuracy of the lower bound $\ubar{C}$ on the compressibility (Eq. \ref{C}). To do so, we consider the upper bound $\bar{C}$ on the compressibility induced by the lower bound $\bar{R}(S)$ on the rate-distortion curve. We find that the lower bound $\ubar{C}$ remains close to the upper bound $\bar{C}$ across almost all networks in Supporting Table 1 (Supporting Fig. \ref{error}\textit{C}). For comparison, Zachary's karate club exhibits a difference $\bar{C} - \ubar{C}$ that is typical for the networks in Supporting Table 1 (Supporting Fig. \ref{error}\textit{C, right}). Interestingly, among the five outliers with an abnormally large difference $\bar{C} - \ubar{C}$, four are the road networks in Supporting Table 1 (Supporting Fig. \ref{error}\textit{C, right}). As discussed in the main text, the abnormal compression properties of road networks likely reflects the unique physical constraints on their structure. Together, the results of this section verify that the upper bound $\bar{R}(S)$ on the rate-distortion curve and the corresponding lower bound $\ubar{C}$ on the compressibility provide reasonable estimates of the true information properties of the real networks analyzed in the main text.

\section{Compressing model networks}
\label{rand_nets}

In our investigations of network compressibility, we analyzed a number of model networks, including Erd\"{o}s-R\'{e}nyi, $k$-regular (Fig. 3\textit{B} in the main text), stochastic block (Fig. 4\textit{A} in the main text), and scale-free (Fig. 4\textit{F} in the main text) networks. Here, we study the structure of optimal compressions in these model networks, as well as networks with hierarchical structure, wherein small tightly-connected groups of nodes connect to form larger, but looser groups \cite{Ravasz-01}.

\begin{figure*}[t]
\centering
\includegraphics[width = .75\textwidth]{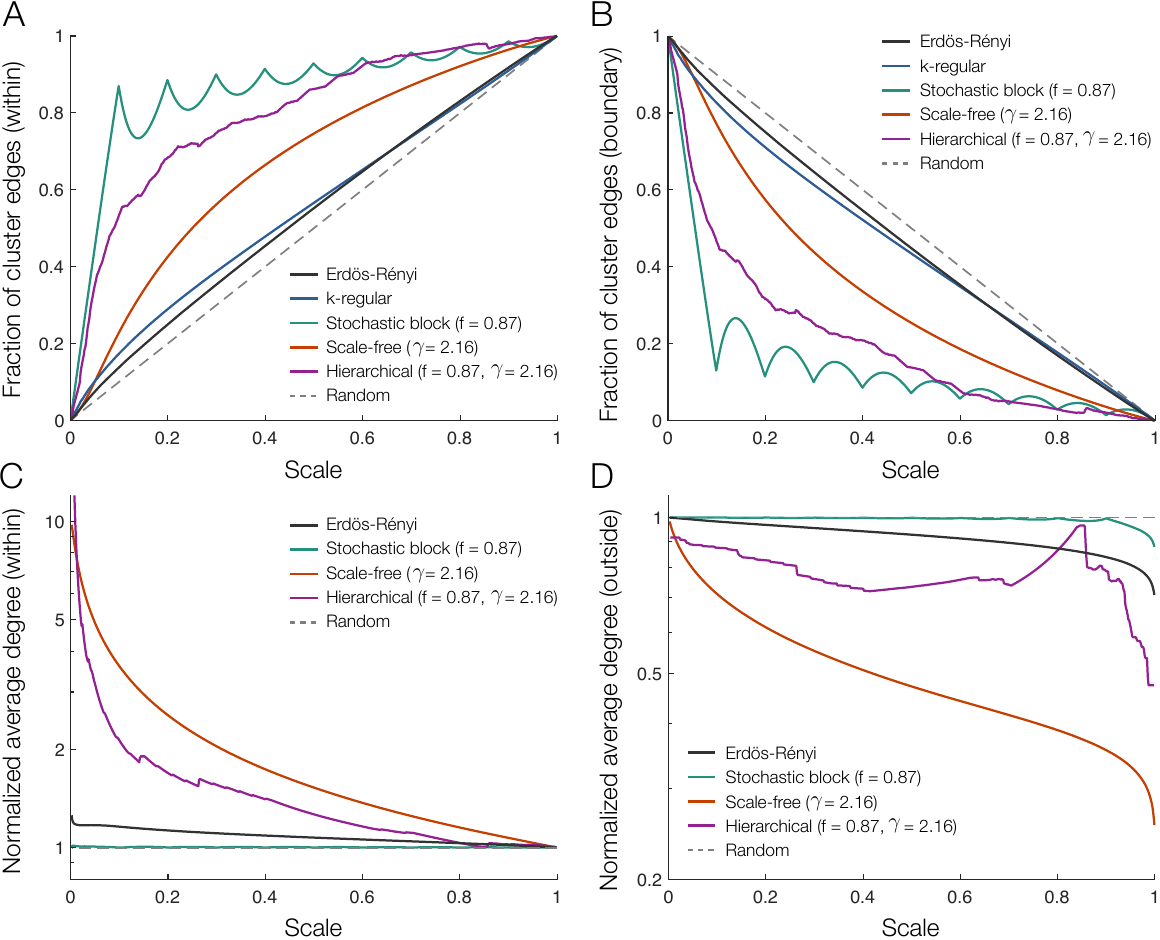}
\captionsetup{labelformat=empty}
\caption{\textbf{Supporting Fig. \ref{rand_cluster}. Structure of optimal clusterings in model networks.} (\textit{A}-\textit{B}) Fraction of the $k_c$ edges emanating from the large cluster that connect to nodes either within the cluster $G_{cc}/k_c$ (\textit{A}) or outside the cluster $1 - G_{cc}/k_c$ (\textit{B}) as a function of the scale $S$. (\textit{C}-\textit{D}) Average degrees of nodes inside (\textit{C}) and outside (\textit{D}) the large cluster, normalized by the average degree of the network, as a function of the scale $S$. In panels (\textit{C}-\textit{D}), data are not displayed for $k$-regular networks because the normalized degree of all nodes is one by definition. Across all panels, the data for Erd\"{o}s-R\'{e}nyi (black), $k$-regular (blue), stochastic block (10 modules; green), and scale-free (red) networks represent averages over 50 randomly-generated networks, each with $N = 10^3$ nodes and average degree $\left< k\right> = 100$. The hierarchical data (magenta) are computed using a Ravasz-Barab\'{a}si network \cite{Ravasz-01} with four recursive levels ($N = 625$, $\left< k\right> = 6.32$). The fraction $f = 0.87$ of within-module edges in the stochastic block networks and the exponent $\gamma = 2.16$ for the scale-free networks are chosen to match the hierarchical network.\label{rand_cluster}}
\end{figure*}

To recall, based on the information content of different edges (see Sec. \ref{edges}), we predicted that optimal compressions would tend to maximize the number of edges within the one large cluster and minimize the number of edges on the boundary and outside of the cluster. In the main text, we confirmed these predictions for the real networks in Supporting Table 1 (see Fig. 2 in the main text). In Supporting Fig. \ref{rand_cluster}, we verify that these results extend to all of the model networks listed above.

First, across all model networks considered, we find that the edges emanating from the one large cluster tend to connect to nodes within the cluster (Supporting Fig. \ref{rand_cluster}\textit{A}) and avoid crossing over to nodes outside of the cluster (Supporting Fig. \ref{rand_cluster}\textit{B}). Interestingly, we observe stark differences in the properties of optimal compressions between different network models. For example, although Erd\"{o}s-R\'{e}nyi and $k$-regular networks do not contain large-scale structure, optimal compressions are still able to identify groups of nodes that are slightly more strongly connected to each other than to the rest of the network. Meanwhile, optimal compressions in stochastic block, scale-free, and hierarchical networks identify groups of nodes with much stronger within-group connectivity. In fact, for stochastic block networks, it is clear that optimal compressions identify the built-in modules, with the one large cluster iteratively enveloping each module one by one as the scale increases (Supporting Fig. \ref{rand_cluster}\textit{A-B}).

Second, for Erd\"{o}s-R\'{e}nyi, scale-free, and hierarchical networks, we confirm that optimal compressions select groups of nodes with higher degrees than average (Supporting Fig. \ref{rand_cluster}\textit{C}) while omitting low-degree nodes (Supporting Fig. \ref{rand_cluster}\textit{D}). As expected, this effect is much stronger in scale-free and hierarchical networks (which exhibit large heterogeneities in node degree) than in Erd\"{o}s-R\'{e}nyi networks (which do not contain large-scale structure). We remark that in $k$-regular networks, there are no differences in node degrees by definition. Similarly, for stochastic block networks, we find a negligible difference between the degrees of nodes inside versus outside the large cluster. This final result once again suggests that optimal compressions in stochastic block networks focus on their modular, rather than heterogeneous, structure. Together, the results of this section demonstrate that the structure of optimal compressions observed in real networks (Fig. 2 in the main text) extends to a range of model networks. Moreover, for hierarchical networks, we establish that optimal compressions cluster together groups of nodes that are more tightly connected than scale-free networks (Supporting Fig. \ref{rand_cluster}\textit{A}), yet higher in degree than stochastic block networks (Supporting Fig. \ref{rand_cluster}\textit{C}), thereby capitalizing on both the modular and heterogeneous properties of hierarchical organization.

\section{Directed networks}
\label{directed}

In the main text, in order to develop analytic predictions for the structure of optimal clusterings and the impact of network structure on compressibility, we focused on the special case of undirected networks. Specifically, we predicted (and confirmed) that optimal compressions would tend to maximize the number of edges within the one large cluster and minimize the number of edges on the boundary and outside of the cluster (Fig. 2\textit{D}-\textit{E} in the main text). In turn, these findings led to the predictions that network compressibility (Eq. \ref{C_supp}) should increase with both modular structure and heavy-tailed degrees, which we confirmed in undirected real and model networks (Fig. 4 in the main text). Here we demonstrate that these central results from the main text are not limited to undirected networks, but in fact generalize to directed networks.

\subsection*{Structure of optimal compressions}

Our clustering algorithm described in the main text is general, applying to any weighted, directed network. We can therefore use the algorithm to compute optimal compressions and rate-distortion curves $\bar{R}(S)$ for the directed real networks listed in Supporting Table 1. We remark that among the 69 real networks studied in the main text, 38 have directed versions (see Supporting Table 1), which we analyze here.

As was the case for undirected networks (Fig. 2\textit{B} in the main text), we find that optimal compressions in directed networks tend to form one large cluster with the maximum possible size $N-n+1 = SN$, where $S = 1- \frac{n-1}{N}$ is the scale of description, and $n-1$ minimal clusters containing one node each (Supporting Fig. \ref{clustering}\textit{A}). To analyze the structure of the large cluster, in the main text we divided the edges in an undirected network into three categories (Fig. 2\textit{C} in the main text): those within the cluster, those outside of the cluster, and those on the cluster boundary. For directed networks, we can further divide boundary edges into two categories -- those connecting from inside to outside the cluster and those connecting from outside to inside the cluster -- thereby resulting in a total of four edge types (Supporting Fig. \ref{clustering}\textit{B}).

\begin{figure*}[t!]
\centering
\includegraphics[width = .9\textwidth]{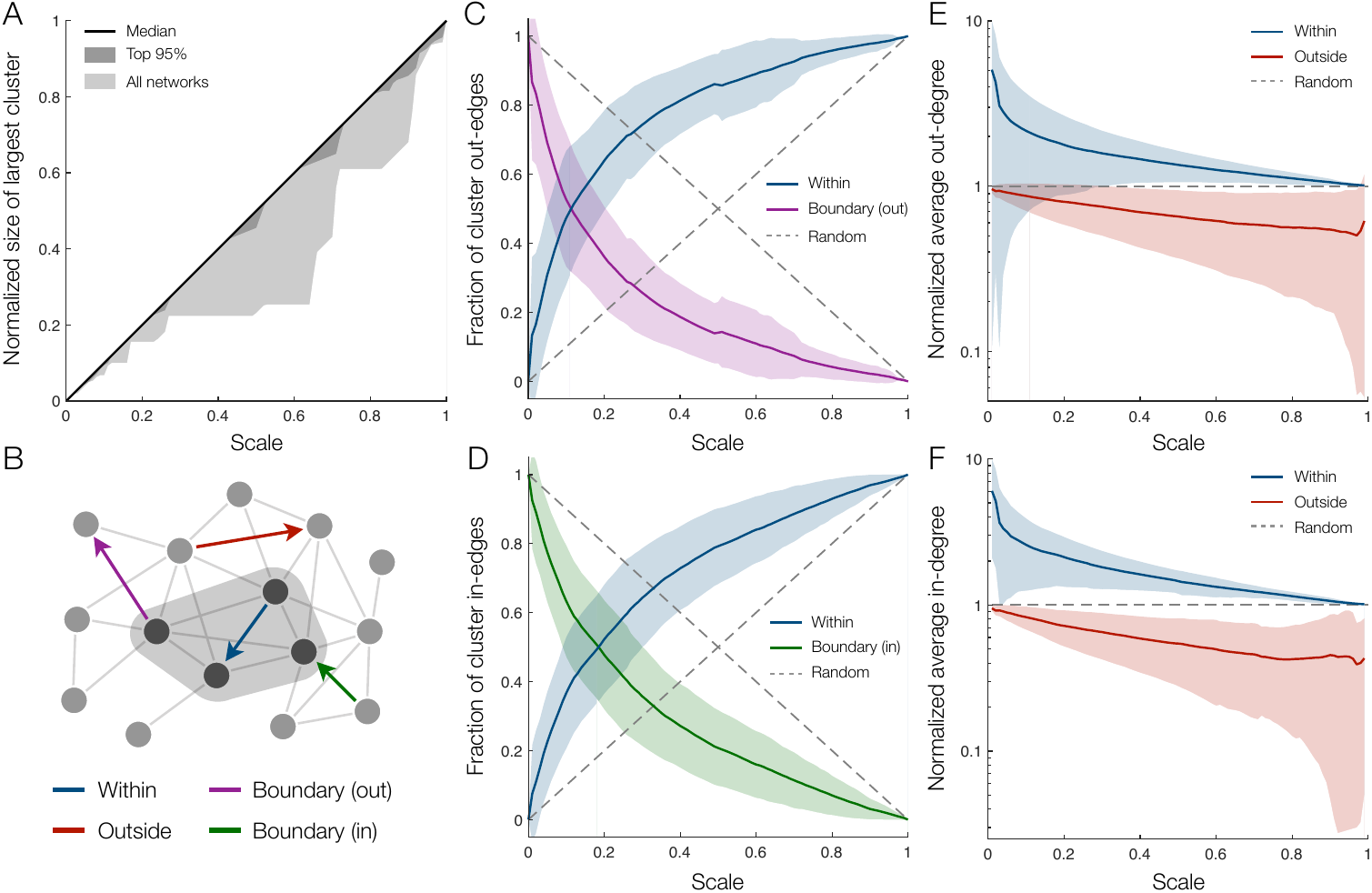}
\captionsetup{labelformat=empty}
\caption{\textbf{Supporting Fig. \ref{clustering}. Structure of optimal clusterings in directed networks.} (\textit{A}) Size of the largest cluster in a compression, normalized by the size of the network $N$, as a function of the scale $S$ for the directed real networks in Supporting Table 1. The median over real networks (solid line) matches the largest possible normalized cluster size, $(N - n + 1)/N = S$. (\textit{B}) Illustration of edges within the one large cluster (blue), outside the cluster (red), crossing the boundary out of the cluster (purple), and crossing the boundary into the cluster (green). (\textit{C}) Fraction of the $k_c^{\text{out}}$ edges emanating from the large cluster that either connect to nodes outside the cluster $1 - G_{cc}/k_c^{\text{out}}$ (purple) or remain within the cluster $G_{cc}/k_c^{\text{out}}$ (blue) as a function of the scale $S$. (\textit{D}) Fraction of the $k_c^{\text{in}}$ edges incident to the large cluster that emanate either from nodes outside the cluster $1 - G_{cc}/k_c^{\text{in}}$ (green) or within the cluster $G_{cc}/k_c^{\text{in}}$ (blue) as a function of the scale $S$. (\textit{E}-\textit{F}) Average out-degree (\textit{E}) and in-degree (\textit{F}) of nodes inside (blue) and outside (red) the large cluster, normalized by the average degree of the network, as a function of the scale $S$. In panels (\textit{C}-\textit{F}), solid lines and shaded regions represent averages and one-standard-deviation error bars, respectively, over the directed real networks in Supporting Table 1, and dashed lines correspond to clusters with nodes selected at random. \label{clustering}}
\end{figure*}

For undirected networks, in the main text we demonstrated that the one large cluster had tight within-cluster connectivity (maximizing the number of edges inside the cluster) and sparse connectivity to the rest of the network (minimizing the number of boundary edges; Fig. 2\textit{D} in the main text). Indeed, this result generalizes to directed networks, with the one large cluster favoring within-cluster edges over both outgoing (Supporting Fig. \ref{clustering}\textit{C}) and incoming (Supporting Fig. \ref{clustering}\textit{D}) boundary edges. In the main text, we also demonstrated that the one large cluster sought to include high-degree nodes and exclude low-degree nodes in undirected networks (Fig. 2\textit{E} in the main text). Here, we confirm that this result also applies to directed networks, with the one large cluster containing nodes with both larger out-degrees (Supporting Fig. \ref{clustering}\textit{E}) and in-degrees (Supporting Fig. \ref{clustering}\textit{F}) than the rest of the network. Together, these results establish that our predictions about the structure of optimal compressions (see Sec. \ref{edges}) generalize to directed networks.

\subsection*{Impact of network structure on compressibility}

We are now prepared to study the compressibility of directed networks. In the main text, we demonstrated that the compressibility of undirected networks increases with the logarithm of the average degree (Fig. 3\textit{E} in the main text). Moreover, based on the structure of optimal clusterings (Fig. 2 in the main text), we hypothesized (and confirmed) that the compressibility of undirected networks increases with both transitivity and degree heterogeneity (Fig. 4 in the main text). Here, we demonstrate that each of these results about the impact of network structure on compressibility generalizes to directed networks.

We begin by analyzing the dependence of network compressibility on average degree. For a directed network with adjacency matrix $G$, where $G_{ij} = 1$ if there is a directed edge from node $i$ to node $j$, the out-(in-)degree of node $i$ is given by $k_i^{\text{out}} = \sum_j G_{ij}$ ($k_i^{\text{in}} = \sum_j G_{ji}$). Despite each node possibly having different out- and in-degrees, the average out- and in-degrees in a network are equal, since
\begin{equation}
\left<k^{\text{out}}\right> = \frac{1}{N}\sum_i k^{\text{out}}_i = \frac{1}{N}\sum_{ij} G_{ij} = \frac{1}{N}\sum_j k^{\text{in}}_j = \left<k^{\text{in}}\right>.
\end{equation}
Therefore, even for directed networks, we can simply discuss the average degree. In Supporting Fig. \ref{structure}\textit{A}, we demonstrate that the compressibility $\ubar{C}$ (Eq. \ref{C_supp}) of real directed networks grows logarithmically with the average degree, confirming that our prediction for undirected networks (Fig. 3\textit{D}-\textit{E} in the main text) extends to directed networks.

\begin{figure*}[t!]
\centering
\includegraphics[width = .75\textwidth]{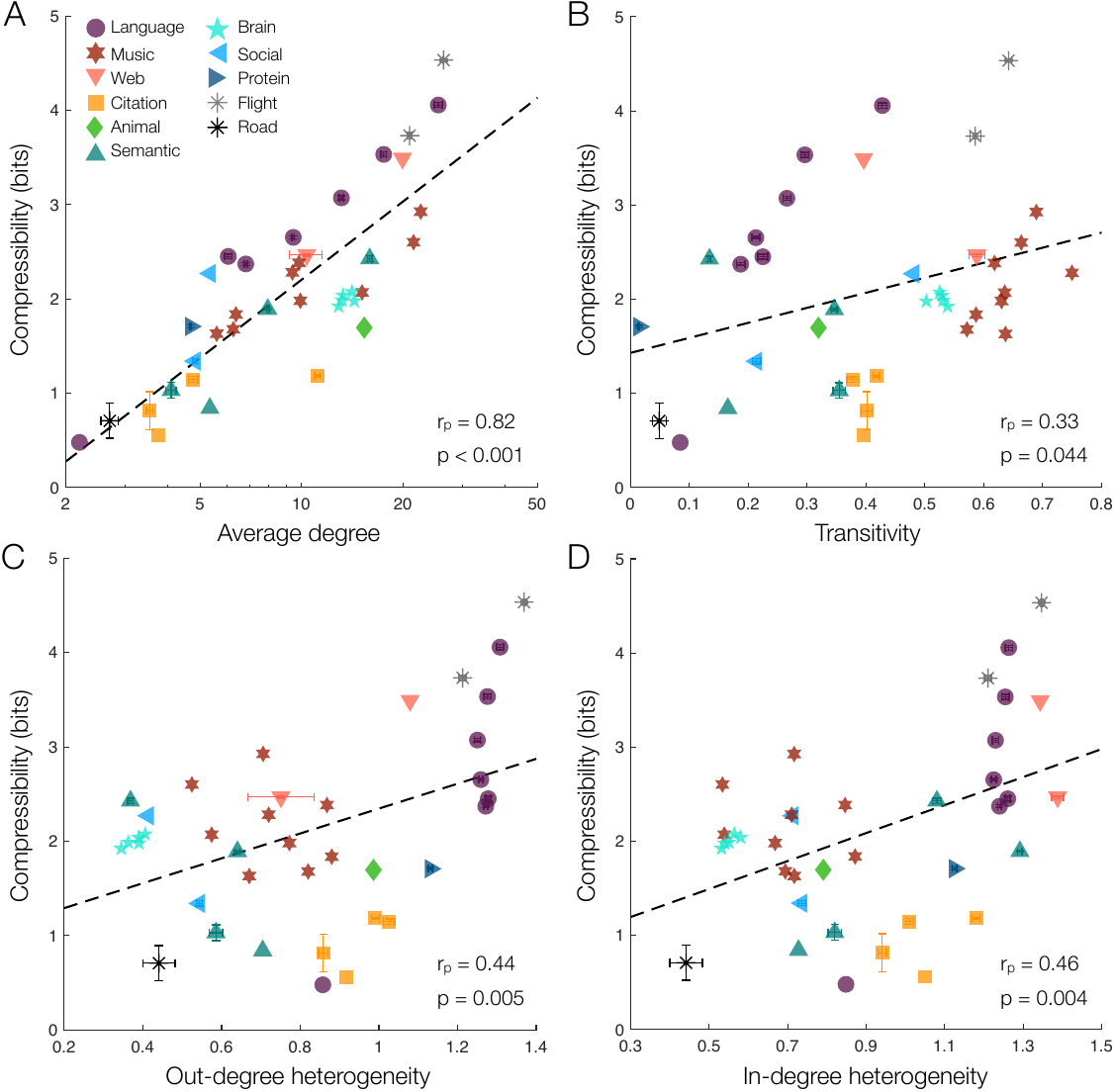}
\captionsetup{labelformat=empty}
\caption{\textbf{Supporting Fig. \ref{structure}. Compressibility of directed networks.} (\textit{A}) Compressibility $\ubar{C}$ (Eq. \ref{C_supp}) versus average degree for the directed real networks in Supporting Table 1. We note that average degree is plotted on a log scale. The dashed line indicates a logarithmic fit. (\textit{B}) Compressibility $\ubar{C}$ versus transitivity (quantified by the average clustering coefficient) for directed real networks (Supporting Table 1) with a linear best fit (dashed line). (\textit{C}-\textit{D}) Compressibility $\ubar{C}$ versus out-degree heterogeneity (\textit{C}) and in-degree heterogeneity (\textit{D}) for directed real networks (Supporting Table 1) with linear best fits (dashed lines). In all panels, for networks of size $N > 10^3$, the data points and error bars represent means and standard deviations over 50 randomly-sampled subnetworks of $10^3$ nodes each (see Materials and Methods in the main text).}
\label{structure}
\end{figure*}

We now consider the impact of transitivity on the compressibility of directed networks. As in the main text, we quantify transitivity using the average clustering coefficient. For undirected networks, the clustering coefficient of a node $i$ is the number of edges connecting the neighbors of $i$ divided by the ${k_i \choose 2} = k_i(k_i-1)/2$ possible connections, and the average clustering coefficient is given by averaging over the nodes in a network. Although the clustering coefficient was originally defined for undirected networks, this definition has since been extended to directed networks \cite{Fagiolo-01}. In a directed network, the clustering coefficient of a node $i$ is the number of connected pairs among its $k_i^{\text{out}} + k_i^{\text{in}}$ out- and in-neighbors divided by the $k_i^{\text{out}} + k_i^{\text{in}} \choose 2$ possible connections. Just as we found for undirected networks (Fig. 4\textit{A}-\textit{E} in the main text), the compressibility of directed networks is significantly correlated with the average clustering coefficient (Supporting Fig. \ref{structure}\textit{B}), thereby indicating that transitivity and the presence of tightly-knit communities serve to make networks more compressible.

Finally, we study the impact of heterogeneous (or heavy-tailed) degrees on the compressibility of directed networks. For directed networks, there are two definitions of degree heterogeneity: that of out-degrees $\left<k_i^{\text{out}} - k_j^{\text{out}}\right>/\left< k^{\text{out}}\right>$ and that of in-degrees $\left<k_i^{\text{in}} - k_j^{\text{in}}\right>/\left< k^{\text{in}}\right>$. We find that the compressibility of directed networks is significantly correlated with both the out-degree (Supporting Fig. \ref{structure}\textit{C}) and in-degree (Supporting Fig. \ref{structure}\textit{D}) heterogeneities. Thus, even in directed networks, heavy-tailed degree distributions with well-connected hubs serve to increase network compressibility. Together, these results demonstrate that the central conclusions from the main text generalize to directed networks; namely, that strong transitivity and degree heterogeneity -- the two defining features of hierarchical organization \cite{Ravasz-01} -- increase the compressibility of complex networks.

\section{Robustness to clustering heuristics}
\label{robust}

\begin{figure*}[t!]
\centering
\includegraphics[width = \textwidth]{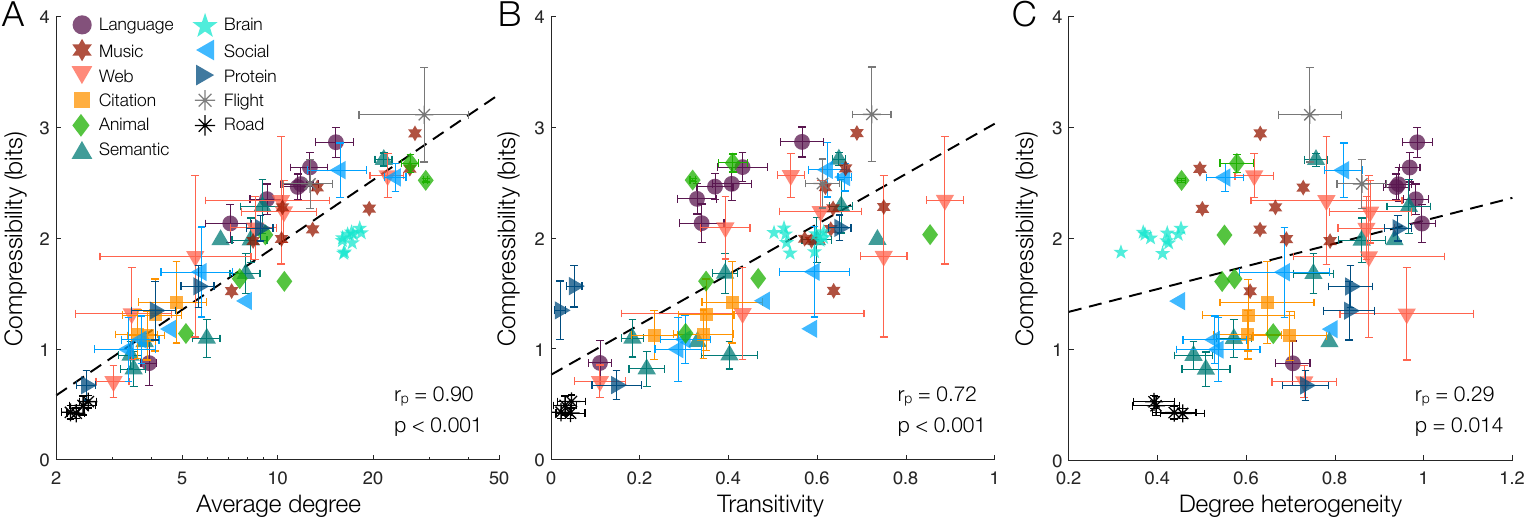} \\
\raggedright
\captionsetup{labelformat=empty}
\caption{\textbf{Supporting Fig. \ref{structure_all}. Compressibility without clustering heuristics.} Compressibility $\ubar{C}$ versus average degree (\textit{A}), transitivity (\textit{B}), and degree heterogeneity (\textit{C}) for the real networks in Supporting Table 1. Compressibility is computed using the brute-force clustering algorithm that tests all pairs of clusters at each iteration rather than a reduced number of pairs selected through heuristics (see Sec. \ref{speedup}). In all panels, for networks of size $N > 100$ data points and error bars represent means and standard deviations over 50 randomly-sampled subnetworks of $100$ nodes each (see Materials and Methods in the main text).}
\label{structure_all}
\end{figure*}

In the main text, we computed rate-distortion curves $\bar{R}(S)$ using our clustering algorithm with the pair-selection heuristics described in Sec. \ref{speedup}. We found that network compressibility increased with average degree (Fig. 3\textit{D-E} in the main text), transitivity (quantified by average clustering coefficient; Fig. 4\textit{D-E} in the main text), and degree heterogeneity (Fig. 4\textit{I-J} in the main text). Here, we confirm that these results are not simply due to our choices of clustering heuristics. To do so, we recompute the compressibilities $\ubar{C}$ of the networks in Supporting Table 1 using the brute-force implementation of our clustering algorithm that searches through all pairs of clusters at each iteration (see Sec. \ref{speedup}). We remark that, in order to employ the brute-force implementation, for each network of size $N > 100$ we analyze 50 randomly-sampled subnetworks of 100 nodes each. Using the brute-force algorithm, we confirm that the compressibility increases with average degree (Supporting Fig. \ref{structure_all}\textit{A}), transitivity (Supporting Fig. \ref{structure_all}\textit{B}), and degree heterogeneity (Supporting Fig. \ref{structure_all}\textit{C}). These observations demonstrate that the central results from the main text are robust to our choice of pair-selection heuristic in the clustering algorithm.

\section{Network datasets and processing}
\label{datasets}

The real-world networks analyzed in the main text are listed and briefly described in Supporting Table 1. The web, citation, animal, semantic, social, protein, flight, and road networks are gathered from online network repositories. The language and music networks were generated by the authors previously (see below) \cite{Lynn-07}. The brain networks are generated from structural and functional data gathered and analyzed previously (see below)\cite{Van-01, Bertolero-02}.

For the language networks, we developed code to (i) remove punctuation and white space, (ii) filter words by their part of speech, and (iii) record the transitions between the filtered words \cite{Lynn-07}. Here we focus on networks of transitions between nouns, noting that the same methods can be used to record transitions between other parts of speech. The raw text was gathered from Project Gutenberg (\texttt{gutenberg.org/wiki/Main\_Page}).

\addtocounter{figure}{-1}
\begin{figure*}
\centering
{\huge
\begin{tabular}{l l l l l l}
\hline
\textbf{Type} & Name & $N$ & $E$ & Description & Reference \\
\hline
\hline
\textcolor{MyMaroon}{\textbf{Language}} & Shakespeare: Combined works$^{*}$ & 11,234 & 97,892 & Noun transitions. & \cite{Shakespeare-01} \\
& Homer: Iliad$^{*}$ & 3,556 & 23,608 & Noun transitions. & \cite{Homer-01} \\
& Plato: Republic$^{*}$ & 2,271 & 9,796 & Noun transitions. & \cite{Plato-01} \\
& Jane Austen: Pride and Prejudice$^{*}$ & 1,994 & 12,120 & Noun transitions. &  \cite{Austen-01} \\
& William Blake: Songs of Innocence...$^{*}$ & 370 & 781 & Noun transitions. & \cite{Blake-01} \\
& Miguel de Cervantes: Don Quixote$^{*}$ & 6,090 & 43,682 & Noun transitions. & \cite{Cervantes-01} \\
& Walt Whitman: Leaves of Grass$^{*}$ & 4,791 & 16,526 & Noun transitions. & \cite{Whitman-01} \\
\hline
\textcolor{MyRed}{\textbf{Music}} & Michael Jackson: Thriller$^{*}$ & 67 & 446 & Note transitions. & \cite{Jackson-01} \\
& Beatles: Hard Day's Night$^{*}$ & 41 & 212 & Note transitions. & \cite{Beatles-01} \\
& Queen: Bohemian Rhapsody$^{*}$ & 71 & 961 & Note transitions. & \cite{Queen-01} \\
& Toto: Africa$^{*}$ & 39 & 163 & Note transitions. & \cite{Toto-01} \\
& Mozart: Sonata No 11$^{*}$ & 55 & 354 & Note transitions. & \cite{Mozart-01} \\
& Beethoven: Sonata No 23$^{*}$ & 69 & 900 & Note transitions. & \cite{Beethoven-01} \\
& Chopin: Nocturne Op 9-2$^{*}$ & 59 & 303 & Note transitions. & \cite{Chopin-01} \\
& Bach: Clavier Fugue 13$^{*}$ & 40 & 143 & Note transitions. & \cite{Bach-01} \\
& Brahms: Ballade Op 10-1$^{*}$ & 69 & 670 & Note transitions. & \cite{Brahms-01} \\
\hline
\textcolor{MyOrange}{\textbf{Web}} & Google internal$^{*}$ & 12,354 & 142,296 & Hyperlinks between internal Google cites. & \cite{Palla-01, Kunegis-01} \\
& Education & 2,622 & 6,065 & Hyperlinks between education webpages. & \cite{Gleich-01,Rossi-01} \\
& EPA & 2,232 & 6,876 & Hyperlinks between pages linking to www.epa.gov. & \cite{DeNooy-01,Rossi-01} \\
& Indochina & 9,638 & 45,886 & Hyperlinks between pages in Indochina. & \cite{Boldi-01,Rossi-01} \\
& 2004 Election blogs$^{*}$ & 793 & 13,484 & Hyperlinks between blogs on US politics. & \cite{Adamic-01,Kunegis-01} \\
& Spam & 3,796 & 36,404 & Hyperlinks between spam pages. & \cite{Castillo-01, Rossi-01} \\
& WebBase & 6,843 & 16,374 & Hyperlinks gathered by web crawler. & \cite{Boldi-01,Rossi-01} \\
\hline
\textcolor{MyYellow}{\textbf{Citation}} & arXiv Hep-Ph$^{*}$ & 12,711 & 139,500 & Citations in Hep-Ph section of the arXiv. & \cite{Leskovec-01, Kunegis-01} \\
& arXiv Hep-Th$^{*}$ & 7,464 & 115,932 & Citations in Hep-Th section of the arXiv. & \cite{Leskovec-01, Kunegis-01} \\
& Cora$^{*}$ & 3,991 & 16,621 & Citations between scientific papers. & \cite{Subelj-01, Kunegis-01} \\
& DBLP$^{*}$ & 240 & 858 & Citations between scientific papers. & \cite{Ley-01,Kunegis-01} \\
\hline
\textcolor{MyGreen}{\textbf{Animal}} & Dolphins & 62 & 159 & Social relationships. & \cite{Lusseau-01, Kunegis-01} \\
& Little Rock Lake & 183 & 2,434 & Food web of animal consumption. & \cite{Martinez-01, Kunegis-01} \\
& Macaques & 62 & 325 & Dominance relationships. & \cite{Takahata-01, Kunegis-01} \\
& Sheep & 24 & 91 & Dominance relationships. & \cite{Hass-01, Kunegis-01} \\
& Wetlands$^*$ & 128 & 2,075 & Food web of animal consumption. & \cite{Ulanowicz-01, Kunegis-01} \\
& Zebras & 23 & 105 & Social interactions. & \cite{Sundaresan-01, Kunegis-01} \\
\hline
\textcolor{MyTurquoise}{\textbf{Semantic}} & Game of Thrones & 796 & 2,823 & Character co-occurrences. & \cite{Kunegis-01} \\
& Algebra & 278 & 3,553 & Concept co-occurrences. & \cite{Christianson-01} \\
& Bible & 1,707 & 9,059 & Pronoun co-occurrences. & \cite{Kunegis-01} \\
& Les Miserables & 77 & 254 & Character co-occurrences. & \cite{Kunegis-01} \\
& Edinburgh Thesaurus$^{*}$ & 7,754 & 226,518 & Word similarities in human experiments. & \cite{Kiss-01,Batagelj-01} \\
& Roget Thesaurus$^{*}$ & 904 & 3,447 & Linked semantic categories. & \cite{Roget-01, Batagelj-01} \\
& Glossary terms & 60 & 114 & Words used in definitions of other words. & \cite{Batagelj-01} \\
& FOLDOC$^{*}$ & 13,274 & 90,736 & Same as above (computing terms). & \cite{Howe-01, Batagelj-01} \\
& ODLIS$^{*}$ & 1,802 & 12,378 & Same as above (information science terms). & \cite{Reitz-01,Batagelj-01} \\
\hline
\textcolor{MyCyan}{\textbf{Brain}} & Structural connectivity 1$^*$ & 100 & 806 & Structural connections between brain regions. & \cite{Van-01} \\
& Structural connectivity 2$^*$  & 100 & 858 & Structural connections between brain regions. & \cite{Van-01} \\
& Structural connectivity 3$^*$  & 100 & 865 & Structural connections between brain regions. & \cite{Van-01} \\
& Structural connectivity 4$^*$  & 100 & 916 & Structural connections between brain regions. & \cite{Van-01} \\
& Structural connectivity 5$^*$  & 100 & 906 & Structural connections between brain regions. & \cite{Van-01} \\
& Functional connectivity 1 & 100 & 811 & Functional correlations between brain regions. & \cite{Van-01} \\
& Functional connectivity 2 & 100 & 800 & Functional correlations between brain regions. & \cite{Van-01} \\
& Functional connectivity 3 & 100 & 830 & Functional correlations between brain regions. & \cite{Van-01} \\
& Functional connectivity 4 & 100 & 854 & Functional correlations between brain regions. & \cite{Van-01} \\
& Functional connectivity 5 & 100 & 799 & Functional correlations between brain regions. & \cite{Van-01} \\
\hline
\textcolor{MyBlue}{\textbf{Social}} & Zachary's karate club & 34 & 78 & Interactions between karate club members. & \cite{Zachary-01, Kunegis-01} \\
& Facebook & 13,130 & 75,562 & Subset of the Facebook network. & \cite{Viswanath-01, Kunegis-01} \\
& arXiv Astr-Ph & 17,903 & 196,972 & Coauthorships in Astr-Ph section of arXiv. & \cite{Leskovec-01, Kunegis-01} \\
& arXiv Hep-Th & 22,721 & 2,672,975 & Coauthorships in Hep-Th section of arXiv. & \cite{Leskovec-01, Kunegis-01} \\
& Adolescent health$^{*}$ & 2,155 & 8,970 & Friendships between students. & \cite{Moody-01, Kunegis-01} \\
& Highschool$^{*}$ & 67 & 267 & Friendships between highschool students. & \cite{Coleman-01, Kunegis-01} \\
& Jazz & 198 & 2,742 & Collaborations between jazz musicians. & \cite{Gleiser-01, Kunegis-01} \\
\hline
\textcolor{MyNavy}{\textbf{Protein}} & C. elegans & 453 & 2,025 & Interactions between metabolites. & \cite{Duch-01, Kunegis-01} \\
& Human (Figeys) & 2,239 & 6,452 & Protein interactions. & \cite{Ewing-01, Kunegis-01} \\
& Human (Stelzl)$^*$ & 1,706 & 6,207 & Protein interactions. & \cite{Stelzl-01, Kunegis-01} \\
& Yeast & 1,870 & 2,277 & Protein interactions. & \cite{Jeong-01, Kunegis-01} \\
\hline
\textcolor{MyGrey}{\textbf{Flight}} & US Flights$^*$ & 1,574 & 28,236 & Flights between US airports. & \cite{Kunegis-01} \\
& OpenFlights$^*$ & 2,939 & 30,501 & Flights between world cities. & \cite{Opsahl-01, Kunegis-01} \\
\hline
\end{tabular}}
\end{figure*}

\addtocounter{figure}{-1}
\begin{figure*}
\centering
{\huge
\begin{tabular}{l l l l l l}
\hline
\textbf{Type} $\quad\quad$ & Name & $N$ & $E$ & Description & Reference \\
\hline
\hline
\textbf{Road} & EuroRoad & 1,174 & 1,417 & Network of roads between European cities. $\quad\quad\quad\quad\quad$ & \cite{Vsubelj-01, Kunegis-01} \\
& Chicago$^*$ & 12,982 & 39,018 & Network of roads in Chicago. & \cite{Eash-01, Kunegis-01} \\
& New York City $\quad\quad\quad\quad\quad\quad\quad\quad\quad$ & 264,346 & 730,100 & Network of roads in New York. & \cite{Kunegis-01} \\
& Bay Area & 321,270 & 794,830 $\quad$ & Network of roads in San Francisco area. & \cite{Kunegis-01} \\
\hline
\end{tabular}}
\captionsetup{labelformat=empty}
\caption{\textbf{Supporting Table 1. Real networks analyzed in the main text.} For each network we list its type; name and whether it has a directed version (denoted by *); number of nodes $N$; number of edges $E$; brief description; and reference.}
\end{figure*}

For the music networks, we read in audio files in MIDI format using the \texttt{readmidi} function in MATLAB (R2018a). For each song, we split the notes by their channel, which represents the different instruments. For each channel, we created a network of note transitions. We then create a transition network representing the entire song by aggregating the transitions between notes across the different channels. The MIDI files were gathered from \texttt{midiworld.com} and from \texttt{kunstderfuge.com}. Our code and data are available upon request from the corresponding author.

For the brain networks, we study the structural and functional connectivity of five randomly-selected subjects from the Human Connectome Project \cite{Van-01}. For all subjects, the brain is divided into 100 predefined cortical regions \cite{Thomas-01}. The structural connectivity networks reflect physical white-matter tracts between brain regions measured using diffusion tensor imaging (DTI). We threshold the structural networks to only include connections between regions that are stronger than the mean. The functional connectivity networks reflect Pearson correlations between regional brain activity. Specifically, the activity is defined by blood-oxygen-level-dependent (BOLD) functional magnetic resonance imagining (fMRI) signals. To ensure that the functional networks have approximately the same edge density as the structural networks, we threshold the functional networks to only include connections stronger than the mean plus one standard deviation.

% Bibliography
\bibliography{GraphLearningBib}

\end{document}